\documentclass[12pt]{article}
\usepackage{epsfig}
\usepackage{amssymb}
\usepackage{pslatex}

\makeatletter

\@addtoreset{equation}{section}
\makeatother
\newcommand{\bea}{\begin{eqnarray}}
\newcommand{\eea}{\end{eqnarray}}
\newcommand{\be}{\begin{eqnarray}}
\newcommand{\ee}{\end{eqnarray}}

\def\fR{{\mathfrak R}}
\def\fL{{\mathfrak L}}
\def\fQ{{\mathfrak Q}}
\def\fS{{\mathfrak S}}

\def\rmd{{\rm d}}

\begin{document}

\begin{titlepage}
\vskip1cm
\begin{flushright}
SNUST 091101 \\
$\mathbb{UOSTP}$ {\tt 09112}
\end{flushright}
\vskip0.1cm
\centerline{\Large \bf Integrability of ${\cal N}=6$ Chern-Simons Theory }
\vskip0.3cm
\centerline{\Large \bf at}
\vskip0.3cm
\centerline{\Large \bf Six Loops and Beyond}
\vskip1.25cm
\centerline{\large
 Dongsu Bak$^{\,a}$, Hyunsoo Min$^{\,a}$, Soo-Jong Rey$^{\,b,\,c}$
}
\vspace{0.75cm}
\centerline{\sl a) Physics Department,
University of Seoul, Seoul 130-743 {\rm KOREA}}
\vspace{0.25cm}
\centerline{\sl b) School of Physics and Astronomy,
Seoul National University, Seoul 151-742 {\rm KOREA}}
\vspace{0.25cm}
\centerline{\sl c) School of Natural Sciences, Institute
for Advanced Study, Princeton NJ 08540 {\rm USA}}
\vspace{0.50cm}
\centerline{\tt dsbak@uos.ac.kr,  \ \
hsmin@dirac.uos.ac.kr, \ \  sjrey@snu.ac.kr}
\vspace{1cm}
\centerline{ABSTRACT}
\vspace{0.75cm}
\noindent
We study issues concerning perturbative integrability of ${\cal N}=6$ Chern-Simons theory at planar and
weak `t Hooft coupling regime. By Feynman diagrammatics, we derive so called maximal-ranged interactions in the quantum dilatation generator, originating from homogeneous and inhomogeneous diagrams. These diagrams require proper regularization of not only ultraviolet but also infrared divergences. We first consider standard operator mixing method. We show that homogeneous diagrams are obtainable by recursive method to all orders. The method, however, is not easily extendable to inhomogeneous diagrams. We thus consider two-point function method and study both operator contents and spectrum of the quantum dilatation generator up to six loop orders. We show that, of two possible classes of operators, only one linear combination actually contributes. Curiously, this is exactly the same combination as in ${\cal N}=4$ super Yang-Mills theory. We then study spectrum of anomalous dimension up to six loops.  We find that the spectrum agrees perfectly with the prediction based on quantum integrability. In evaluating the six loop diagrams, we utilized remarkable integer-relation algorithm (PSLQ) developed by Ferguson, Baily and Arno.


\end{titlepage}

\section{Introduction}
The AdS/CFT correspondence\cite{Maldacena:1997re} continues revealing remarkable relations
between gauge and gravity theories.
The most extensively studied so far is the correspondence between the
four-dimensional ${\cal N}=4$ superconformal Yang-Mills theory and the Type
IIB superstring theory on AdS$_5 \times \mathbb{S}^5$\cite{Maldacena:1997re}.
Importantly, both theories admit Lagrangian formulation, which involve two
coupling parameters: the rank $N$ of the gauge group $G$ and the `t Hooft
coupling constant $\lambda$ for the former, and the string coupling $g_{\rm s}$
and the curvature scale $R$ (as measured in string unit) for the latter. As such,
perturbatively, one can compute physical observables in both theories in double
series of the respective parameters and test the correspondence by comparing a
given observable extracted from each sides. In the planar limit, $N \rightarrow \infty$
 and $g_{\rm s} \rightarrow 0$, remarkable agreement was discovered between the
two sides for a variety of observables. The agreement is largely attributed to
 the integrability structure\cite{Minahan:2002ve}-\cite{Hofman:2006xt}.

Recently, a Type IIA counterpart was discovered and added to the list of the
 AdS/CFT correspondence.
The ABJ(M) theory is (2+1)-dimensional ${\cal N}=6$ superconformal Chern-Simons
theory and was proposed as the holographic dual to the Type IIA superstring theory
 on AdS$_4 \times \mathbb{CP}^3$\cite{Aharony:2008ug,Aharony:2008gk}.
A question of interest is whether
the two sides in this Type IIA counterpart also have an integrability structure.
Recently, positive indications toward the quantum integrability were
accumulated\cite{classicalintegrability}-\cite{AhnDoreyNepomechie}.
 At strong coupling side,
Lax pair construction of the integrability was shown at leading
order\cite{classicalintegrability}.
 At weak coupling side, there were more indications. At two loops, spin
chain Hamiltonian was computed explicitly for the $SO(6)$ subsector and was
shown integrable\cite{Mina, Bak, Bak2}. See also Ref.~\cite{twoloop}.
At four loops, spectrum of the spin chain Hamiltonian
was shown to agree with the prediction of the
integrability\cite{BakMinRey,Mina2}.

In this paper, we further continue our
previous investigations \cite{Bak, Bak2, BakMinRey} concerning
integrability of the ${\cal N}=6$ Chern-Simons theory in the
weak 't Hooft coupling regime by computing the dilatation operator to six loops.
Given that the integrability is in place up to four loops, why bother six loops?
We contend that there are two important aspects that arise beginning at six loops
and beyond: operator contents and recursive structure of the long-range spin chain.
With these two issues on focus, we shall test the integrability of the ${\cal N}=6$
 Chern-Simons theory at six loop order.
As in \cite{BakMinRey}, we shall focus on magnon excitation in the $SU(2)$ subsector,
compute operator structure {\sl and} spectrum of the spin chain Hamiltonian up to six
loops and check them against the prediction based on the integrability and the
{\sl centrally extended}
$[\mathfrak{psu}(2 \vert 2) \oplus \mathfrak{psu}(2 \vert 2)] \ltimes \mathbb{R}^{2,1}$
superalgebra of excitation symmetry.

The off-shell $\mathfrak{psu}(2|2)$ superalgebra of the excitation symmetry is spanned
 by the two $su(2)$ rotation generators ${\fR}^a\!_b$, ${\fL}^\alpha\!_\beta$, the
 supersymmetry generator ${\fQ}^\alpha_a$ and the superconformal generator ${\fS}^a_\alpha$.
The off-shell configuration is characterized by $sl(2, \mathbb{R})$ central
charges $\mathfrak{C}, \mathfrak{K}, \mathfrak{K}^*$~\cite{Beisert:2005tm}.
Their (anti)commutators are  given by \cite{Beisert:2005tm}
\bea
&& [{\fR}^a\!_b, \,\, \mathfrak{J}^c]= \delta^c_b \,
\mathfrak{J}^a-{1\over 2}
\delta^a_b \, \mathfrak{J}^c\,,\ \ \
[{\fL}^\alpha\!_\beta, \,\, \mathfrak{J}^\gamma]= \delta^\gamma_\beta  \,
\mathfrak{J}^\alpha-{1\over 2}
\delta^\alpha_\beta \, \mathfrak{J}^\gamma\,
\nonumber\\
&& \{ {\fQ}^\alpha_a, \,\, \fS^b_\beta\}= \delta^b_a   {\fL}^\alpha\!_\beta
+ \delta^\alpha_\beta \, {\fR}^b\!_a
+ \delta^b_a  \delta^\alpha_\beta \mathfrak{C}            \nonumber \\
&& \{ {\fQ}^\alpha_a, \,\, \fQ^\beta_b\}=\epsilon^{\alpha\beta}
\epsilon_{ab} \mathfrak{K}\,,\ \ \
\{ {\fS}^a_\alpha, \,\, \fS_\beta^b\}=\epsilon_{\alpha\beta}
\epsilon^{ab} \mathfrak{K}^* \,.
\eea
The central charges $\mathfrak{C}$ is related to the energy
by $E= \mathfrak{C}$, while $\mathfrak{K}, \mathfrak{K}^*$ introduced at off-shell
are related to momentum of the magnon. Acting on a magnon transforming in the fundamental
 representation, closure of the superalgebra leads to the relation among the central charges
\bea
 E^2 = \mathfrak{C}^2 = {1\over 4} + 4 \mathfrak{K} \mathfrak{K}^*.
\eea
The central charges $\mathfrak{K}, \mathfrak{K}^*$ are in turn
related to an excitation momentum $P$. More
generally, a bound-state of $Q$ elementary magnons transforming in
 higher-dimensional representations can be studied. The off-shell analysis
 was sufficient to determine the dispersion relation. It is
\bea
E(P) ={1\over 2}\sqrt{Q^2+ 16 h^2(\lambda) \sin^2{P\over 2}}\,\, , \label{energy}
\eea
where $h(\lambda)$ is a function of the `t Hooft coupling parameter $\lambda$.

Restricting to large $N$ limit and $SO(6)$ sector of the $OSp(6 \vert 4, \mathbb{R}$) superconformal symmetry group, the quantum dilatation operator was computed explicitly at two loops from which an integrable alternating spin chain Hamiltonian and Bethe ansatz equations were identified \cite{Mina,Bak,Bak2}.
Aspects of the integrability were explored further beyond two loops by focusing on diagrams generating {\sl maximal-ranged} terms. These are the diagrams in which interaction vertices range over lattice sites of the spin chain Hamiltonian maximally. In \cite{BakMinRey}, we computed four-loop contribution to these terms and found that the spectrum fits with the prediction based on the integrability and the excitation symmetry.

At each order in perturbation theory, depending on the range the `spin' flavors at different sites are permuted, maximal-ranged terms in the dilatation operator are further classifiable into maximal-shuffling and next-to-maximal-shuffling terms. In ${\cal N}=4$ SYM theory, it was found by Gross, Mikhailov and Roiban \cite{GMR} that maximal-shuffling terms are computable recursively. Inspection of relevant Feynman diagrams indicates that all diagrams contributing to maximal-shuffling terms are related by a recursion relation and hence resummable to an exact all-order result. On the other hand, diagrams contributing to nonmaximal-shuffling terms are combinatorially too complicated to be resummable.
One might anticipate that a similar argument exists for the ${\cal N}=6$ ABJ(M) theory since conformal interactions are tightly constrained by large amount of supersymmetry. Indeed, we shall find that the maximal-shuffling terms in the dilatation operator originates from the same class of skeleton diagrams which we call {\sl homogeneous} diagrams. We were able to perform all-order resummation and show that they match with the structure of dilatation operator predicted by integrability. The nonmaximal-shuffling terms receive contribution from another class of skeleton diagrams which we call {\sl inhomogeneous} diagrams. As these diagrams are not recursively resummable and afflicted with potential infrared divergences, we need to resort to an alternative approach for direct evaluation. In the second half of work, we thus adopt two-point function method and compute six-loop contribution to maximal-ranged interactions from both homogeneous and inhomogeneous diagrams. This method amounts in dual Type IIA string theory to deriving time evolution Hamiltonian of a single non-interacting string.

Key results of our work point to the followings. The dilatation operator at six loops are extractable free of infrared divergences from two-point correlation functions of gauge invariant operators, which was already utilized in our earlier study at four loops \cite{BakMinRey}. Moreover, the maximal-ranged interactions are consistent with the integrability and that, curiously, operator contents of the long-range spin chain Hamiltonian is identical to those of the ${\cal N}=4$ super Yang-Mills theory, viz. Inozemtsev spin chain~\cite{Inozemtsev:2002vb, Serban:2004jf}.

This paper is organized as follows.
We begin with description of the expected spectrum based on the integrability
and  prediction for the maximal shuffling coefficients to all orders.
In section 3, we recapitulate all-loop computation of maximal shuffling terms
in ${\cal N}=4$ super Yang-Mills theory, obtained first by Gross, Mikhailov and Roiban
in \cite{GMR}. We then extend the method to ${\cal N}=6$ Chern-Simons theory
and find two classes of diagrams contributing to the maximal shuffling.
The first class of diagrams, homogeneous diagrams, is computable by a
straightforward extension of the method in \cite{GMR} and yields a result exactly
parallel to the ${\cal N}=4$ super Yang-Mills theory. The second class,
 inhomogeneous diagrams, is not computable by the method of \cite{GMR}
or variants of it: these diagrams are afflicted with infrared divergences.
We conclude that all-order derivation for the maximal shuffling part of the dilatation
operator is not straightforward in ${\cal N}=6$ Chern-Simons theory.
We thus resort to computing operator contents and spectrum of their
anomalous dimensions order by order in perturbation theory. In section 4,
we study 6-loop contribution to the anomalous dimension directly defined by
two point correlation functions of operators.
In section 5, we identify  the matrix structures of
the homogeneous and the inhomogeneous
maximal-ranged interactions at 6-loops.
In section 6, we compute the maximal-ranged part of the 6-loop
Hamiltonian and confirm
the prediction based on the integrability.  We also show that
only one particular choice arises  among two possible
maximal shuffling operators. This operator is the same
as the one arising in the  ${\cal N}=4$ super Yang-Mills theory.
In section 7, we extend our results to parity non-conserving ${\cal N}=6$
ABJ theory, whose gauge group is U($M)\times$U$(N)$ with $ N < M < N+k$.
The last section is devoted to the concluding remarks. In the appendices,
we relegate several technical details. Appendix A illustrate a comparative calculation of inhomogeneous term in operator-mixing method. Appendix B presents several lattice operator
identities. Appendix C discusses derivation of skeleton 4-loop diagrams. Appendix D
explains implementation of numerical integration and the remarkable integer relation
PSLQ algorithm.

\section{Hamiltonian and Spectrum From Integrability}

We begin with consequence of the quantum integrability and the off-shell superalgebra to
the spin chain Hamiltonian of a sub-sector of our interest. Consider single-trace bosonic operators in the
${\cal N}=6$ ABJM theory of the type 
\bea
\Psi\,[I_1\, I_2 \, I_3 \, I_4 \, I_5 \,\cdots I_{2L}]={\rm tr}[ \, Y_{I_1}\,
Y^\dagger_{I_2}\,Y_{I_3}\, Y^\dagger_{I_4}\, Y_{I_5}\,
\cdots\,Y^\dagger_{I_{2L}}]\,.
\label{singletrace}
\ee
Here, $2L$ is the total number of the sites. We shall take the infinite volume limit $L \rightarrow \infty$
and view a particular ordering of the operator as a lattice spin chain wave function $\Psi$. Gauge invariant operators place at odd sites the $Y_I\,\, (I,J=1,2,3,4)$ elementary scalar fields transforming as ${\bf 4}$ under the $SU(4)$ R-symmetry and at even sites the $Y^\dagger_I$ conjugate fields. Under the $SU(4)\simeq SO(6)$ R-symmetry, these fields transform as ${\bf 4}$ and $\overline{\bf 4}$, respectively.
We denote $\{Y_1, Y_2, Y_3, Y_4\}$ as $\{A_1, A_2, B^\dagger_1, B^\dagger_2\}$
where $A_a, B_{\dot{a}}$ ($a,\dot{a}=1,2$)
transform under the $SU(2)_A$ and the SU(2)$_B$ subgroups of $SU(4)$.
We then consider a subset of the operators (\ref{singletrace}), where  we put only  $A_a\,\,$/$\,\,B_{\dot{a}}$ fields at the odd$\,\,$/$\,\,$even sites, respectively.
Explicitly, they are the following restricted set
of operators 
\be
\Psi\,[\,a_1 \,{\dot{a}}_2\, a_3\, \dot{a}_4\, {a}_5
\,\cdots {\dot{a}}_{2L}]={\rm tr}[\, A_{a_1}\,
B_{\dot{a}_2}\,A_{a_3}\, B_{\dot{a}_4}\, A_{a_5}\,\cdots\,
 B_{\dot{a}_{2L}}\,]\,.
\label{seto}
\ee
Since the only possible interaction between $Y_I$ and $Y^{\dagger J}$ fields is the contraction
$\delta_I^{J}$ group theoretically,
there cannot be any interaction between ${\bf 4}$ and $\overline{\bf 4}$ representations acting on the above type of states.
Therefore, for this restricted set of operators, the odd-site chain
and even-site chain are decoupled from each other. Thus, there will be
$A$-type magnon and $B$-type magnon propagating independently without any interactions between them.
For unrestricted operators, there are interactions between them, but the above choice avoids unnecessary complexity in investigating the integrability.
For the reference vacuum, we take the ferromagnetic state 
$|0\rangle =
\Psi\,[\,1\,1\,1\,1\,1\,\cdots \,\,1]=
{\rm tr}[\, A_{1}\,
B_{1}\,A_{1}\, B_{1}\, A_{1}\, \cdots\, B_{1}\,]
$.
In \cite{BakMinRey}, we explained how elementary excitations above the reference vacuum are
organized by the centrally extended $[psu(2|2)\oplus psu(2|2)] \ltimes \mathbb{R}^{2,1}$ superalgebra.
The first $psu(2|2)$ acts on the flavors of the magnons formed by exciting odd sites
of the spin chain, while the second $psu(2 \vert 2)$ acts on the magnons at even sites.
The bosonic $su(2)$ subalgebra of $psu(2|2)$ superalgebra corresponds to exciting
$Y_2, \,\,Y_4$ for the odd sites and $Y^\dagger_2, \,\,Y^\dagger_4$
for the even sites. In \cite{BakMinRey}, we carried out careful study of the representations of the centrally extended superalgebra for the asymptotic spin chain where $2L$ is sent to infinity. The quantum integrability of the dilatation operator implies the factorization of multi-magnon S-matrices into product of two-magnon S-matrices satisfying the Yang-Baxter equations~\footnote{The all-loop proposal of the  Bethe ansatz and S-matrices of ${\cal N}=6$ ABJM theory was put
forward in Ref.~\cite{allloop}.}. For both the ABJM and the ABJ theories, we also showed \cite{Bak, Bak2} that dynamics of the magnons on even sites and on odd sites are governed by two separate transfer matrices $\tau^{\rm alt}(u, \gamma), \overline{\tau}^{\rm alt}(v, - \gamma)$ for arbitrary spectral parameters $u, v, \gamma $ and that, using the Yang-Baxter equations, they are mutually commuting
\bea
[\tau^{\rm alt}(u, \gamma), \tau^{\rm alt}(v, -\gamma)] = 0, \qquad [\overline{\tau}^{\rm alt}(u, \gamma), \overline{\tau}^{\rm alt}(v, - \gamma)]=0, \qquad [\tau^{\rm alt}(u, \gamma), \overline{\tau}^{\rm alt}(v, - \gamma)] = 0 \ .
\eea
Their moments are
\bea
{\cal Q}_n = \partial_u^{n-1} \ln \tau^{\rm alt} (u, \gamma) \Big\vert_{u=0} \qquad \mbox{and}
\qquad
\overline{\cal Q}_n = \partial_v^{n-1} \ln \overline{\tau}^{\rm alt} (v, - \gamma) \Big\vert_{v=0} \hskip0.5cm
(n=1,2,\cdots)
\label{commuting}
\eea
of which ${\cal Q}_2 + \overline{\cal Q}_2$ is proportional to the dilatation operator. They all depend on
the spectral parameter $\gamma$ -- a nonzero value of $\gamma$ is an indication that the dilatation operator
and all other higher moments are not invariant under lattice parity transformation. We found in \cite{Bak, Bak2} that $\gamma = 0$ not only for the ABJM theory but surprisingly also for the ABJ theory, which is parity non-conserving. From (\ref{commuting}), it followed that there are two sets of mutually commuting, infinitely many conserved charges
\bea
[{\cal Q}_m, {\cal Q}_n]=0, \qquad
[\overline{\cal Q}_m, \overline{\cal Q}_n] = 0, \qquad
[{\cal Q}_m, \overline{\cal Q}_n]=0 \ .
\eea
It was argued \cite{AhnDoreyNepomechie} that these mutually commuting conserved charges are responsible for reflectionless property of the S-matrix elements between a magnon on even sites and a magnon on odd sites.

Invariance of the S-matrices under the off-shell superalgebra transformations was crucial to
fix the representation as well as the spectrum of elementary magnons. The analysis (recapitulated in the previous section) shows that the dispersion relation of an elementary magnon takes the form (\ref{energy})
with $Q=1$ and $P$ is pseudo-momentum of the magnon and $h(\lambda)$ is an interpolating function of the 't Hooft coupling $\lambda$.
In addition, the pseudo-momentum $P$ that specifies the central charge and the magnon spectrum can also
be a function of the lattice momentum $p$ defined by translation in the spin chain. The functional form of the interpolating function $h(\lambda)$ and the pseudo-momentum $P(p)$ are not determinable by the symmetry alone and require extra inputs of explicit computations either of ${\cal N}=6$ Chern-Simons theory at weak coupling or of string worldsheet sigma-model at strong coupling. In the previous work \cite{BakMinRey}, we found that $P(p)=p$ holds up to four-loop order. In this work, we shall assume this as an input and proceed for computation of six-loops and beyond.

Perturbatively, the interpolating function $h^2(\lambda)$ is expandable as
\be
h^2(\lambda)=
\lambda^2\left(1 +\sum^\infty_{\ell=1} h_{2\ell} \,\, \lambda^{2\ell}\right) 
= \lambda^2 \left( 1 + h_2\,\, \lambda^2 + h_4\,\, \lambda^4+ \cdots\right)
\label{hlambda}
\ee
where, for the leading term, we use the result of the two-loop
computation in \cite{Mina,Bak}. Recently, in \cite{Mina2}, the next coefficient $h_2$ was computed to be
$ h_2=4 \zeta(2)-16$. 
Thus, in terms of the lattice momentum $p$, the magnon dispersion relation can be
expanded as
\bea
E(p) &=& \sum^\infty_{n=0}
\lambda^{2n}\sum^{n}_{l=0} e_{2n,2l} \sin^{2l} {p\over 2}
\nonumber\\
&=&  \left({1\over 2}\right) +
\Bigl(4\,\, \sin^2 {p\over 2}\Bigr) \lambda^2   +
\Bigl( 4 h_2\,\, \sin^2 {p\over 2} - 16 \,\sin^4 {p\over 2}\Bigr)
\lambda^4
\nonumber\\
&+&
\Bigl( e_{6,2} \sin^2 {p\over 2} +e_{6,4} \sin^4 {p\over 2}
+e_{6,6} \sin^6 {p\over 2}
\Bigr)
\lambda^6 +
\cdots \ .
\label{spectruma}
\eea
Note that $e_{0,0}=1/2$ corresponds to the classical scaling dimension of the elementary scalar fields
$Y_2, Y_4, Y_2^\dagger, Y_4^\dagger$ and $e_{2n,0}=0 \,\,\,(n\,\, >\,\, 0)$ is required by one-third of the ${\cal N}=6$ supersymmetry preserved by the ferromagnetic
vacuum state.~\footnote{In Chern-Simons theory,
the choice of regularization method is known to be a subtle issue. Detailed study in \cite{Bak, Bak2} utilized the dimensional reduction and obtained $e_{2,0} = 0$ at two loops. This confirms that the dimension reduction is a gauge invariant and supersymmetric regulator at least up to two loop order. Whether
the corresponding Ward identities are satisfied at higher loops is an open problem that needs to be checked.
All higher loop diagrams involving gauge and ghost fields are afflicted by the problem but diagrams involving matter fields only are not. Our previous \cite{BakMinRey} and the present works deal only with Feynman diagrams of the latter type and hence are free from this open problem.}
The two-loop coefficient $e_{2,2}=4$ was found by explicit computation in \cite{Mina, Bak}.  The four-loop coefficient $e_{4,4}=-16$ was computed in \cite{BakMinRey}, while $e_{4,2}= 4 h_2$ was argued in \cite{Mina2}.

On the other hand the expected spectrum based on the quantum
integrability and the off-shell superalgebra representation theory is expandable as
\bea
E(p) &=& {1\over 2}\sqrt{1+ 16\, h^2(\lambda) \sin^2 {p\over 2} }\nonumber\\
&=&
{1\over 2} +   \left(4 \sin^2 {p\over 2}\right) \lambda^2 +
\left(
4 \sin^2 {p\over 2}\right) \left[ h_2  -4\, \sin^2
{p\over 2} \right] \lambda^4
\nonumber\\
&+& \left(4 \sin^2 {p\over 2}\right)
  \left[\
 h_4
 - 8 h_2\,\, \sin^2 {p\over 2} + 32\,\, \sin^4 {p\over 2}\ \right]\,\, \lambda^6
+\cdots\,,
\label{conjecture}
\eea
where we use the expansion for $h^2(\lambda)$ in (\ref{hlambda}).
Note that the coefficients $e_{2n,2n}$ are fixed completely by the dispersion relation:
\be
e_{2n,2n}= (-1)^{n+1}\, 4^{n}\,\, {\big(2n-2\big)!\over (n-1)!\,\,n!}\,.
\ee
It is the coefficient of the term $(\sin^2 {p\over 2})^n=(2-e^{ip}-e^{-ip})^n/4^n$.
Therefore, at order $\lambda^{2n}$ of the perturbation theory,
the coefficient of the eigenvalue $e^{\pm i\,p\,n}$ is uniquely fixed as
\be
{\cal C}_{2n} = -
{(2n-2)!\over (n-1)!\,\,n!}. \label{coeff}
\ee
As we shall explain below graphically, $e^{\pm i\,p\,n}$ is the eigenvalue of the maximal-shuffling
operator on the spin chain lattice. From this argument, we conclude that
the coefficients of the maximal-shuffling operators are fixed
by the assumption of the integrability and the representation theory of off-shell symmetry superalgebra. In this work, we shall explicitly compute these coefficients at six loops and compare with (\ref{coeff}). It
constitutes a nontrivial check of the quantum integrability of ${\cal N}=6$ Chern-Simons theory.

To derive the spectrum, following \cite{BakMinRey}, we use the lattice-momentum eigenstates of elementary
magnon. For the A-spin magnon propagating on the odd-site chain, we have
\be
| p\rangle_A = \sum_{\ell=0}^{L-1} e^{i \, \ell p}\, |\dots (2\ell+1)_{A_2}\dots\rangle \qquad \mbox{and} \qquad
| p\rangle_B = \sum_{\ell=1}^{L} e^{i \, \ell p}\, |\dots (2\ell)_{B_2}\dots\rangle\, .
\label{pa}
\ee
Here, $|\dots (2\ell+1)_{A_2}\dots\rangle$ refers that we put $A_2$ at the $2\ell +1$'s site
while we put $A_1$ for the remaining odd-sites, and similarly for the even-site chain. 
Hence, for an elementary magnon state, we may consider two kinds of states
$| p\rangle_A \otimes | 0\rangle_B$ and $| 0\rangle_A \otimes | p\rangle_B$.
Below we shall focus on the odd-site chain as the odd-site and even-site chains behave independently
for the above subset of magnon states we are interested in.

The corresponding integrable Hamiltonian at each order is
well known for the above set of states\footnote{All order generalization to the full spin chain that takes account of ${\bf 4}\!-\!\overline{\bf 4}$ interactions would be extremely interesting.}.
 Since the structure of the even chain is identical to the
odd chain, it is sufficient to focus on the odd-site chain only.
 The zeroth order spin chain Hamiltonian counts the classical scaling dimension of the spins:
\be
H_0 = {1\over 2} \sum^L_{l=0}
\,\, \mathbb{I} \,.
\ee
The two loop part of the Hamiltonian is
given by~\cite{Mina,Bak}
\be
 H_2= 4\sum^{L-1}_{\ell=0}\,\, \mathbb{O}^{2\ell}_{2,2} \qquad
\mbox{where} \qquad
\mathbb{O}^{2\ell}_{2,2}={1\over 4}\,\big[\, \mathbb{I}
-\mathbb{P}_{{2\ell+1},\, {2\ell+3}}\,\big]\,
\ee
and $\mathbb{P}$ is the permutation operator defined by
$\mathbb{P}^{cd}_{ab}=\delta_a^d\,\,\delta_b^c$.
We shall take the infinite volume limit, $L \rightarrow \infty$, and do not consider wrapping interactions.
The corresponding 4-loop Hamiltonian can be identified as
\be
H_4= e_{4,2} \sum^{L-1}_{\ell=0}
\mathbb{O}^{2\ell}_{2,2}
+ e_{4,4}\sum^{L-1}_{\ell=0}
\mathbb{O}^{2\ell}_{4,4}
\quad \mbox{where} \quad
\mathbb{O}^{2\ell}_{4,4}={1\over 16}
\Big[4(\mathbb{I}-\mathbb{P}_{{2\ell+1}, {2\ell+3}}) -(\mathbb{I}-\mathbb{P}_{{2\ell+1}, {2\ell+5}})
\Big].
\nonumber
\ee
This parallels the analysis of the ${\cal N}=4$ super Yang-Mills
theory~\cite{
Beisert:2003tq,Beisert:2003ys,Eden:2004ua}.
The 6-loop Hamiltonian can also be identified as
\bea
H_6 =
{e_{6,2}} \, \sum^{L-1}_{\ell=0}
\, \mathbb{O}^{2\ell}_{2,2}
+ e_{6,4}\, \sum^{L-1}_{\ell=0}
\, \mathbb{O}^{2\ell}_{4,4}
+ e_{6,6}\, \sum^{L-1}_{\ell=0}
\, \Big[
(1-\kappa_6)\,\mathbb{O}^{2\ell}_{6,6}+\kappa_6\,
\widetilde{\mathbb{O}}^{2\ell}_{6,6}\Big]\,.
\eea
Here, $\kappa_6$ is an arbitrary coefficient. We thus see that a new feature arises beginning at six loop order. Up to four loops, candidate spin chain operators consistent with the quantum integrability are uniquely fixed. At  6-loops, operators consistent with the integrability are no longer unique: there are two
commuting, mutually independent operators
\bea
\mathbb{O}^{2\ell}_{6,6}
&=&{1\over 64}\Big[\,\,\mathbb{P}_{{2\ell+1}, {2\ell+5}}\,
\mathbb{P}_{{2\ell+3}, {2\ell+7}}
-\mathbb{P}_{{2\ell+1}, {2\ell+7}} \,\mathbb{P}_{{2\ell+3}, {2\ell+5}}\,
  \nonumber\\
&&\ \ \ \ \ \  +\,\,\,
4\,\mathbb{P}_{{2\ell+1},
{2\ell+5}}-14\, \mathbb{P}_{{2\ell+1}, {2\ell+3}}+10\,\,\mathbb{I}\,\,\,
\Big]\, , \nonumber \\
\widetilde{\mathbb{O}}^{2\ell}_{6,6}
&=&
{1\over 64}\,\Big[\,(\,\mathbb{I}-\mathbb{P}_{{2\ell+1}, {2\ell+7}})
-6\,(\,\mathbb{I}-\mathbb{P}_{{2\ell+1}, {2\ell+5}})
+15\, (\,\mathbb{I}-\mathbb{P}_{{2\ell+1}, {2\ell+3}})\,\,
\Big]\, .
\eea
Notice that these operator contents are identifiable with the Inozemtsev spin
chain system~\cite{Inozemtsev:2002vb, Serban:2004jf}.
Acting on the momentum eigenstate $\vert p \rangle_A$ in (\ref{pa}), these operators are diagonalized with the {\sl same} eigenvalue
\be
\sum^{L-1}_{\ell=0}\mathbb{O}^{2\ell}_{2n,2n}\,\,\,
| p\rangle_A =
\sum^{L-1}_{\ell=0}\widetilde{\mathbb{O}}^{2\ell}_{2n,2n}
\,\,\,| p\rangle_A=
\sin^{2n} {p\over 2} \,\,\, | p\rangle_A\,.
\ee
So, the operator content of the six-loop Hamiltonian is not determinable uniquely by probing a single
magnon dispersion relation. On the other hand, the operators $\mathbb{O}_{6,6}$ and
$\widetilde{\mathbb{O}}_{6,6}$ are distinguishable by acting on a state containing two or more magnon
excitations. Note that both of them include the maximal-shuffling operators that produce the eigenvalues $e^{\pm i\,np}$. From the purely integrability point of view of the long-ranged Heisenberg spin chain,
both operators are allowed and fit perfectly. Below, we shall approach this issue by computing relevant
Feynman diagrams explicitly. We then determine the coefficient $\kappa_6$ and, from it, the operator contents of the spin chain Hamiltonian.

\section{Recursive Method for Maximal-Ranged Interactions}

We begin in this section with a recursive method, first put forward
by Gross, Mikhailov and Roiban \cite{GMR}, of extracting maximal-shuffling terms among 
maximal-ranged interactions. We first redo the computation in ${\cal N}=4$
super Yang-Mills theory and then repeat the computation in ${\cal N}=6$ Chern-Simons theory.

Of maximal-ranged interactions, we are particularly interested in the coefficients 
${\cal C}_{2n}$ of the maximal shuffling operators --- they will provide a direct 
test of the integrability. In ${\cal N}=4$ super Yang-Mills theory, Feynman
diagrams contributing to maximal-shuffling term are easily identifiable.
As will be briefly reviewed below, at each order in perturbation theory, 
there is only one type diagram contributing to maximal-shuffling term. We refer to it as {\sl 
homogeneous} diagram. For example, at 3-loop order, this maximal-shuffling
term is responsible for the operator $\mathbb{O}_{6,6}$. There are also 
non-maximal-shuffling terms generated from maximal-ranged diagrams.

For ${\cal N}=6$ Chern-Simons theory, despite different interaction structures, we find a strikingly
similar pattern repeated. Among the maximal-ranged interactions, maximal-shuffling term arises from 
homogeneous diagram. It turns out this diagram gives rise to the operator $\mathbb{O}_{6,6}$, as in the situation of the ${\cal N}=6$ super Yang-Mills theory. We shall further confirm the coefficients ${\cal C}_{2n}$ of the maximal shuffling terms. Among the maximal-ranged interactions, there are 
also inhomogeneous diagrams that give rise to nonmaximal-shuffled terms. At six-loop order, for instance, these diagrams are responsible for the operator $\kappa_6\,\, \big[\,\widetilde{\mathbb{O}}_{6,6}-
\mathbb{O}_{6,6}\,\big]$. However, by a direct computation, we shall find the coefficient $\kappa_6$ vanishes identically. This is interesting since, a priori, this operator could be generated given that interactions in the ${\cal N}=6$ Chern-Simons theory are different from those in the ${\cal N}=4$ super Yang-Mills theory, this operator could be generated.

\subsection{${\cal N}=4$ super Yang-Mills theory}
We first rederive the all-loop, maximal ranged interactions in ${\cal N}=4$ super
Yang-Mills theory obtained by Gross, Mikhailov and Roiban \cite{GMR}, emphasizing
aspects directly relevant for similar considerations in ${\cal N}=6$ Chern-Simons
theory.

\begin{figure}[ht!]
\centering \epsfysize=6cm
\includegraphics[scale=0.65]{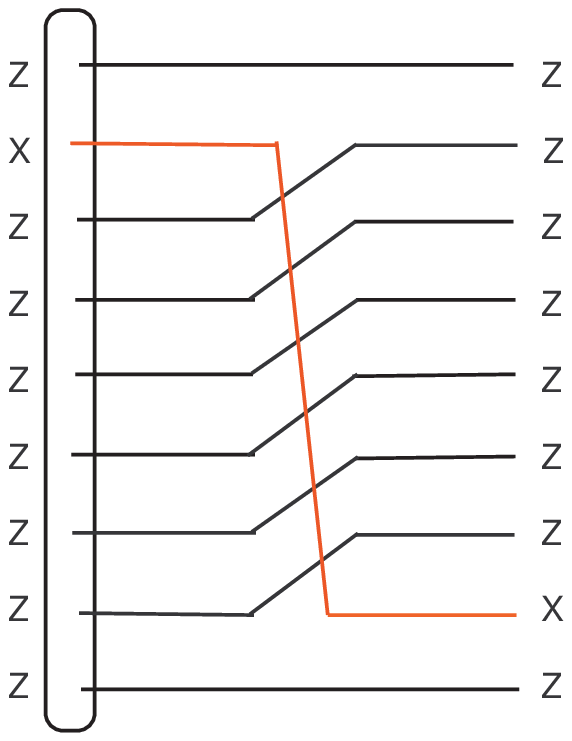}
\caption{\small \sl Homogeneous diagram in maximal-ranged interactions in ${\cal N}=4$ super Yang-Mills theory. It generate maximal-shuffling term in the dilatation operator.}
\label{fig1-1}
\end{figure}

Consider in the ${\cal N}=4$ super-Yang-Mills theory renormalization of composite operators.
Introduce in the theory the dimensional regularization parameter $\epsilon$, $d=4-2 \epsilon$
of ultraviolet divergences. To avoid infrared divergences, a set of external momenta $q$ are
also injected to the operator.
Denote by $I_{\rm bare} (\epsilon, q)$ a collection of regularized Feynman diagrams that
contribute to the renormalization of composite operators of the type (\ref{singletrace}).
Omitting wave function renormalization part proportional to the identity part,
multiplicative renormalizability of the operators asserts that
\be
I_{\rm ren}(q) = {\rm exp}\Big[\,\,{1\over 2\epsilon}\int^1_0
{\rmd t\over t}\ \Big(H_4 (\lambda t)-1\Big)
\,\,\Big]\,\,
I_{\rm bare} (\epsilon, q)
\label{reno1}
\ee
ought to be  finite in the limit where
$\epsilon$ goes to zero.
The generator $H_4 (\lambda)$ is the quantum dilatation operator
generating renormalzation group transformation and
its eigenvalue corresponds to the magnon dispersion relation in the spin chain
interpretation. In perturbation theory,
\bea
I_{\rm bare} (\epsilon, q) = 1 + \sum_{\ell = 1}^\infty I^{(\ell)}_{\rm bare} (\epsilon, q) \ .
\eea

From the $\epsilon$-independence of $I_{\rm ren}(k)$, we get the relation
\be
\int^1_0 {\rmd t\over t}\Big(H_4 (\lambda t)-1\Big) =
-\lim_{\epsilon\rightarrow 0}\,\, 2\epsilon\, \,
{\rm ln}\,\, I_{\rm bare} (\epsilon, q)\,,
\ee
for the operators which do not include the identity part. In planar ${\cal N}=4$
super Yang-Mills theory, the maximal-ranged interactions are generated only by quartic
scalar interactions. Homogeneous diagrams among them are depicted in Fig.~\ref{fig1-1}. To avoid
infrared divergences, we inject a finite momentum
$q$ to the $X$ field from the right in Fig.~\ref{fig1-1}.
It suffices to keep zero momentum for all other $Z$ fields.
The $\ell$-loop contribution to the maximal shuffling diagrams
can be evaluated recursively from the $(\ell-1)$-loop contribution~\cite{GMR}:
\be
I^{(\ell)}_{\rm bare} (\epsilon, q)=\left({4\pi\over q^2}\right)^{\ell\epsilon}
\,\, {1\over \epsilon^\ell \,\, \ell!}\,\,\,
{\left[\,\,\Gamma(1-\epsilon)\,\,\right]^{\ell+1}\,\, \Gamma(1+\ell\epsilon)\over
\Gamma(2-(\ell+1)\epsilon)\,\, \prod^{\ell}_{j=2}\,\big(1-{j\,\epsilon}\big)
}\,\, 
{\hat{a}}^\ell 
\ ,
\ee
where $\hat{a}$ denotes $\lambda\,\, (e^{ip}+ e^{-ip}-2) /(16\pi^2)$
and $p$ denotes the lattice momentum such that $e^{\pm i p}$ generates shift one
lattice site to the left or right in the spin chain. From the integrability,
we expect the Hamiltonian to be
\be
H_4 (\lambda, p) = \sqrt{1-4\,{\hat{a}}} \ =\ \sqrt{1-{\lambda \over
4\pi^2}\,\,(e^{ip}+ e^{-ip}-2)}\,. \label{expected}
\ee

One can check finiteness of the renormalized diagrams $I_{\rm ren}(q)$ in
(\ref{reno1}) order by order in $\lambda$. For instance, we checked this explicitly to the order
$O(\lambda^5)$ using the Mathematica and found that the renormalized diagrams to this order are
indeed finite, canceling all $\epsilon^{-n}\,\, (n > 0)$ singular terms.
As was done in \cite{GMR}, we now show asymptotically that the singular terms of the
regularized amplitude $I_{\rm bare} (\epsilon)$ are canceled by the expected Hamiltonian (\ref{expected})
in the renormalization factor
\be
{\rm exp}\Big[\,\,{1\over 2\epsilon}\int^1_0 {\rmd t\over t}\big(\sqrt{1-4\,\hat{a}\, t}-1\big)
\,\,\Big]\,\, =
{\rm exp}\Big[\,\,{1\over \epsilon}\,\,\Big(\sqrt{1-4\hat{a}}-1+\ln\,\,{2\over
1+\sqrt{1-4\hat{a}}}\,\,\Big)
\,\,\Big]\
\label{scaling}
\ee
in the limit $\epsilon$ goes to zero. To show this, we take $\epsilon\rightarrow 0$ while holding
$x := \ell \,\epsilon$ finite and sum the all-loop contribution by the Euler-McLaughlin formula:
\bea
I_{\rm bare} (\epsilon, q)&=& \sum_{\ell = 0}^\infty I^{(\ell)}_{\rm bare} (\epsilon, q) \nonumber \\
& \simeq &\!
 {1\over\sqrt{\epsilon}}\,\int_0^\infty {\rmd x }\, f_4(x, q)\,
{\rm exp}\Big[\,{1\over \epsilon}\,\Big(\,x\,\big(
\ln\,\hat{a}-\ln\, x\,+2\,\big) +(1-x)\,\ln\,(1-x)
\,\Big)\Big] \ .
\eea
We have relegated all sub-leading remainder to $f_4(x,q)$:
\bea
f_4(x, q) := {1 \over \sqrt{x}} \ e^{-\psi(1)\,x} \ \left({4 \pi \over q^2} \right)^x {\Gamma(1 + x) \over \Gamma (2 - x)} \ ,
\eea
where $\psi(z)$ is the poly-gamma function.
The integral can be evaluated by the saddle-point approximation. At the saddle-point
\be
x_0 ={1\over 2}\,\, (1-\sqrt{1-4\hat{a}})\, ,
\ee
the integral is evaluated as
\be
I_{\rm bare} (\epsilon, q)= {\rm exp}\Big[\,\,-\,{1\over \epsilon}\,\,
\Big(\sqrt{1-4\,\hat{a}}-1+\ln\,\,{2\over
1+\sqrt{1-4\,\hat{a}}}\,\,\Big) + R_4(q)
\,\,\Big]\, .
\ee
The $\epsilon^{-1}$ pole term is precisely inverse of the renormalization factor (\ref{scaling})
dictated by the integrability. As it should be, the renormalization factor is independent of the
infrared regularizing momentum $q$ -- the dependence resides in the
finite remainder function $R_4(q)$.

\begin{figure}[ht!]
\centering \epsfysize=6cm
\includegraphics[scale=0.65]{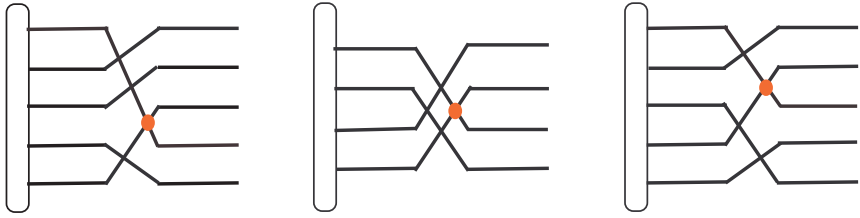}
\caption{\small \sl Inhomogeneous diagram in maximal-ranged interactions in ${\cal N}=4$ super Yang-Mills theory. It generate nonmaximal-shuffling term in the dilatation operator. The circles mark presence of vertices none of whose legs are connected to the operator.}
\label{fig1-11}
\end{figure}

Among the maximal-ranged interactions, there are also inhomogeneous diagrams. At four-loop order, they
were studied in \cite{beisert3loop}. These diagrams are distinguished from homogeneous diagrams that some of the scalar quartic vertices do not connect to the
operator at all. They are responsible for generating nonmaximal-shuffling terms in the dilatation operator. One can convince that these diagrams proliferate rapidly at higher orders in perturbation theory and, even worse, do not show any recursive pattern. Therefore, their contribution needs to be computed individually.
In section 6, adopting two-point function method, we will find that the inhomogeneous diagrams can be 
computed without ambiguity.

\subsection{${\cal N}=6$ Chern-Simons theory}
We now extend the recursive method to ${\cal N}=6$ Chern-Simons theory. One easily see that relevant Feynman diagrams are classifiable again into {\sl homogeneous} and {\sl inhomogeneous}
diagrams. The homogeneous diagrams are planar irreducible diagrams all of whose interaction vertices are
connected to the operator by two internal lines. The {\sl inhomogeneous} diagrams: planar irreducible diagrams some of whose interaction vertices are connected to the operator by one internal line or not connected to the operator at all.

Here, we first consider the homogeneous diagrams as they are easier to evaluate.
\begin{figure}[ht!]
\centering \epsfysize=6cm
\includegraphics[scale=0.8]{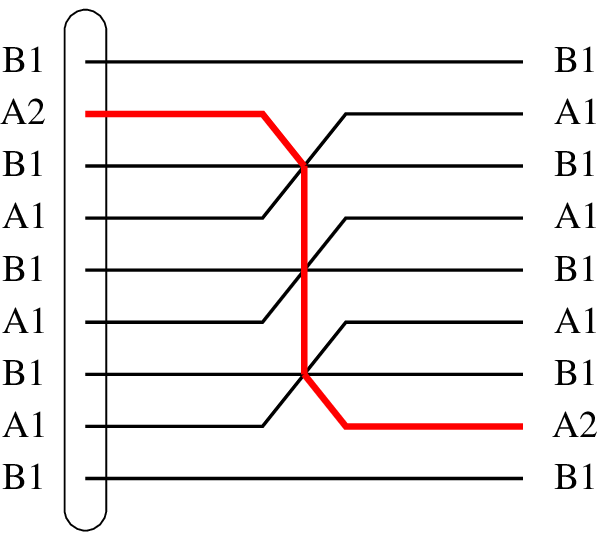}
\caption{\small \sl `homogeneous' maximal shuffling diagrams}
\label{fig1-2}
\end{figure}
The homogeneous diagrams include all maximal shuffling terms.
let us first state the expected scaling of operators via the
anomalous dimension.

Here, as explained in detail in \cite{Bak, Bak2}, we adopt dimension reduction
$2\omega=3-\epsilon$ for regularizing ultraviolet divergences in Feynman diagrams.
To avoid infrared divergences, we again inject momenta $q$ to the composite
operator. Multiplicative renormalizability of composite operator asserts that
\be
{\cal I}_{\rm ren} (q) = {\rm exp}\Big[\,\,{1\over 2\epsilon}
\int^1_0 {dt\over t}\Big(H_3(\lambda^2 t)-{1\over 2}\Big)
\,\,\Big]\,\,
{\cal I}_{\rm bare} (\epsilon, q)
\label{reno2}
\ee
is independent of the regulator $\epsilon$. Recall that, in ${\cal N}=6$
Chern-Simons theory, ultraviolet divergences arise only at even loops. So, in perturbation theory,
\be
{\cal I}_{\rm bare} (\epsilon, q) =1+ \sum^\infty_{\ell =1}\, {\cal I}^{(2 \ell)}_{\rm bare} (\epsilon, q)\,,
\ee
where ${\cal I}^{2\ell}_{\rm bare} (\epsilon, q)$ denotes the $2\ell$-loop regularized diagrams
of order $O(\lambda^{2\ell})$. From $\epsilon$-independence of $I_{\rm ren}(q)$, we get the relation
\be
\int^1_0 {\rmd t\over t}\Big(H_3 (\lambda^2\, t)-{1\over 2}\Big)
= -\lim_{\epsilon\rightarrow 0}\,\,
2\epsilon \,
\ln \, {\cal I}_{\rm bare} (\epsilon, q)\,
\label{H-relation-CS}
\ee
for the operators which do not include the identity part.

As we explained in the previous section, from the integrability, we expect that the Hamiltonian
for the maximal shuffling term is given by
\be
H_{3}(\lambda^2)= {1\over 2}\,\,\sqrt{1-4 \,\hat{b}}\, .
\ee
Here, $\hat{b}=\lambda^2\,\, \big(e^{ip}+e^{-ip}-2\big)$ and $e^{\pm ip}$ should now be interpreted as a left/right shift operator by one lattice spacing on even- or odd-sites of the alternating spin chain. Therefore, the integrability asserts that the renormalization factor is given by
\be
{\rm exp}\Big[\,\,{1\over 2\epsilon}\int^1_0 {\rmd t\over t}\big({1 \over 2} \sqrt{1-4\,\hat{b}\, t}-{1 \over 2}\big)
\,\,\Big]\,\, =
{\rm exp}\Big[\,\,{1\over 2\epsilon}\,\,\Big(\sqrt{1-4\hat{b}}-1+\ln\,\,{2\over
1+\sqrt{1-4\hat{b}}}\,\,\Big)
\,\,\Big]\,.
\label{scaling3}
\ee

To check if ultraviolet divergence of the bare diagrams $I_{\rm bare}(\epsilon, q)$ is inverse of
(\ref{scaling3}), we now evaluate the homogeneous diagrams. At elementary $2\ell + 2$-loops, from Fig.~2, we see that the diagram can be evaluated using $\ell$ many skeleton propagators defined by 1-bubble diagram:
\bea
L_2 (k) = {C_2 \over (k^2)^{2 - \omega}} \qquad \mbox{where} \qquad
C_2 = {\Gamma^2(\omega - 1) \Gamma(2 - \omega) \over (4 \pi)^\omega \Gamma(2 \omega -2)} \ ,
\eea
one skeleton propagator defined by 2-bubble diagram:
\bea
L_3 (k) = {C_3 \over (k^2)^{3 - 2 \omega}} \qquad \mbox{where} \qquad
C_3 = {\Gamma^3 (\omega - 1) \Gamma (3 - 2 \omega) \over (4 \pi)^{2 \omega} \Gamma(3 \omega - 3)}
\eea
and using recursively the skeleton 1-loop integral:
\bea
G(a,b) &:=& (4 \pi)^\omega (p^2)^{a + b - \omega} \int
{\rmd^{2 \omega} k \over (2 \pi)^{2 \omega}}
{1\over (k^2)^a ((k+p)^2)^b} \nonumber \\
&=& {\Gamma(a + b - \omega) \over \Gamma(a) \Gamma(b)} {\Gamma(\omega - a)
 \Gamma(\omega - b) \over \Gamma(2 \omega - a - b)}.
\eea
Denote by ${\cal A}_\ell$ the coefficient that skeleton $\ell$-loop
contributes to the permutation $\mathbb{P}$ in the maximal-shuffling term ${\cal O}= 2 \mathbb{I} - 4\mathbb{P}$ in the dilatation operator.
Label $n$-th loop momenta by $k_n$ and inject an external momentum $q$ at the last skeleton vertex to regulate the loop integrals. At other vertices, there is no need to inject external momenta.
Taking vertex and symmetry factors into account, the skeleton 0-loop (which is actually
elementary 2-loop) contribution reads
\bea
{\cal A}_0(q) &=& (- (2 \pi)^2) \cdot (-4) {C_3 \over (q^2)^{3 - 2 \omega}}.
\label{a0} 
\eea
The skeleton 1-loop (which is actually elementary 4-loop) contribution reads
\bea
{\cal A}_1(q) &=& (- (2 \pi)^2) (-4) \int {\rmd^{2 \omega} k_1 \over (2 \pi)^{2 \omega}}
{\cal A}_0(k_1)
{1 \over k_1^2} {C_2 \over ((k_1 - q)^2)^{2 - \omega} } \nonumber \\
&=& (- (2 \pi)^2)^2 (-4)^2 G(4 - 2 \omega, 2 - \omega) {C_2 C_3 \over
(4 \pi)^\omega (q^2)^{2(3 - 2 \omega)} }.
\label{a1}
\eea
%
%
%
The skeleton 2-loop (which is actually elementary 6-loop) contribution reads
\bea
{\cal A}_2(q) &=& (- (2 \pi)^2) (-4) \int {\rmd^{2 \omega} k_2 \over (2 \pi)^{2 \omega}}
{\cal A}_1(k_2) {1 \over k_2^2} {C_2 \over ((k_2 - q)^2)^{2 - \omega}} \nonumber \\
&=& (4 \pi)^6\,\, G(4 - 2 \omega, 2 - \omega)
G(7 - 4 \omega, 2 - \omega) {C_2^2 C_3 \over (4 \pi)^{2 \omega} (q^2)^{3(3 - 2 \omega)}} \ .
\eea
Recursive pattern is evident. The skeleton $\ell$-th loop (note that this is actually
elementary $2 \ell +2$-loop) contribution reads
\bea
{\cal A}_{\ell}(q) &=&
\prod_{n=1}^{\ell} G(1 + n(3 - 2 \omega), 2 - \omega) \cdot
{(4 \pi)^{2\ell+2 }\,\,C_2^\ell C_3 \over (4 \pi)^{\ell \omega} (q^2)^{(\ell+1)(3 - 2 \omega)}} \ .
\eea
Multiplying ${\hat{b}}^{\ell+1}$ to ${\cal A}_\ell$,
we find the regularized $2\ell$-loop amplitude has the expression
\be
{\cal I}^{(2 \ell)}_{\rm bare} (\epsilon, q)=\left({4\pi\over q^2}\right)^{\ell \epsilon}
\,\, {1\over \epsilon^\ell \,\, \ell!}\,\,\,
{\left[\,\,\Gamma\big(
{1\over 2}-{\epsilon\over 2}
\big)\,\,\right]^{2\ell+1}
\,\, \Gamma(1+ \ell \,\epsilon)\over
\Gamma\big({1\over 2}-{2\ell +1\over 2}
\,\mbox{\small ${\epsilon}$}\,
\big)\,\, \prod^{\ell}_{j=1}\,\big({1\over 2}-{2j+1\over 2}
\,\mbox{\small ${\epsilon}$}\,
\big)
}\,\, \Big[{\hat{b}\over 4 \pi}\Big]^\ell \,.
\ee
Using Mathematica, we checked up to 10-loop orders that the renormalized
diagram $I_{\rm ren}(q)$ in (\ref{reno2}) with the $H_3$
dictated by the integrability is indeed finite in the limit where $\epsilon$ goes to zero.

As in the ${\cal N}=4$ super Yang-Mills theory, we can estimate the asymptotic
behavior of the bare diagram $I_{\rm bare}(\epsilon, q)$ in the limit $\ell \rightarrow \infty$,
$\epsilon \rightarrow 0$ while holding $x := \ell \epsilon$ constant.
%
%
%
%
%
%
By the Euler-McLaughlin formula, we have
\be
{\cal I}_{\rm bare} (\epsilon, q)\simeq
 {1\over\sqrt{\epsilon}}\,\int_0^\infty \,{\rmd x }\, f_3(x,q)\,
{\rm exp}\Big[\,{1\over 2\epsilon}\,\Big(\,2x\,\big(
\ln\,\hat{b}-\ln\,2x\,+2\,\big) +(1-2x)\,\ln(1-2x)
\,\Big)\Big],
\ee
where $f_3(x)$ denotes a sub-dominant remainder
\be
f_3 (x,q)={1\over\sqrt{2x}}\, e^{-\,\,\psi
\big({1\over 2}\big) x} \,\,
\left({4\pi\over q^2}\right)^{x}
\,{\Gamma(1+x)\over \Gamma\big({1\over 2}-x\big)}\, .
\ee
We evaluate the integral by the saddle-point approximation. At the saddle point:
\be
x_0 ={1\over 4}\,\, (1-\sqrt{1-4\hat{b}})\,,
\ee
the integral is given by
\be
{\cal I}_{\rm bare} (\epsilon, q)= {\rm exp}\Big[\,\,-\,{1\over 2\epsilon}
\,\,\Big(\sqrt{1-4\,\hat{b}}-1+\ln\,\,{2\over
1+\sqrt{1-4\,\hat{b}}}\,\,\Big)
\, + R_3 (q) \,\Big]\, .
\ee
We see that the $\epsilon^{-1}$ pole term is precisely the inverse of the renormalization factor in (\ref{scaling3}).
It is remarkable that this all-loop agreement between the homogeneous diagrams and the integrability is closely parallel to the situation in the ${\cal N}=4$ super Yang-Mills theory. Because of this, in the next section, we adopt the two-point function method for deriving quantum dilatation operator. 

\begin{figure}[ht!]
\centering \epsfysize=6cm
\includegraphics[scale=0.8]{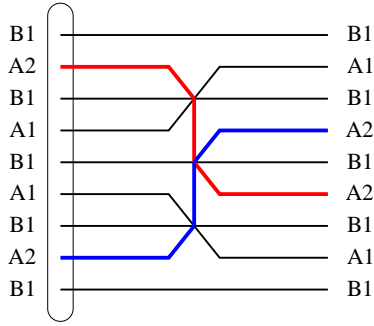}
\caption{\small \sl `inhomogeneous' maximal ranged diagrams}
\label{fig1-3}
\end{figure}

We also need to take account of inhomogeneous diagrams. It is easy to see that this class of diagrams does not include the maximal-shuffling terms. Nevertheless, the interaction range is still maximal, viz. maximal-ranged, at a given order in perturbation theory, as illustrated in Fig.~\ref{fig1-3}. In so far as one just focuses on maximal-shuffling part the spectrum, the homogeneous diagrams are sufficient. If one
would like to identify operator contents of the spin chain Hamiltonian, however, it is
indispensable and crucial to take account of the inhomogeneous diagrams. For instance, at
6-loop order, the inhomogeneous diagrams are responsible for the coefficient $\kappa_6$ of the
operator $\mathbb{O}^{\rm in}_{6,6} =\big[\, \widetilde{\mathbb{O}}_{6,6}-{\mathbb{O}}_{6,6}
\,\big]$.

One would like to see if all-loop contribution of the inhomogeneous diagrams is also obtainable
from the recursive method, much as for the homogeneous diagrams. Here, as in ${\cal N}=4$ super Yang-Mills theory, the relevant diagrams proliferate rapidly at each higher order in perturbation theory and do not exhibit recursive pattern in any obvious way. 

\section{Anomalous dimension matrix}
From now, as in the previous works \cite{Bak, Bak2, BakMinRey},
we shall extract the anomalous dimension matrix of the single-trace operators $\mathbb{O}$ of the type (\ref{singletrace}) from two-point correlation functions:
\be
\langle :\! \mathbb{O}(x)\!: \,\,\, :\! \mathbb{O}(0)\!:\rangle_{\epsilon}
=  {C^{2L}_\epsilon \over
(x^2)^{{1-\epsilon\over 2} \cdot 2L}\,
(x^2)^{\gamma(\epsilon)}
} \ .
\ee
From the dual Type IIA string theory viewpoint, this method amounts to deriving time-evolution Hamiltonian of a single non-interacting string propagating in AdS$_4 \times \mathbb{CP}^3$ spacetime. 
We use the dimensional reduction to regularize ultraviolet divergences.
The two-point correlation functions are related to Feynman loop diagrams $A_{2 \ell}$ by
\bea
 \langle :\!\mathbb{O}(x)\!: \,\,\, :\!\mathbb{O}
(0)\!:\rangle_{\epsilon}
&&= (I_\epsilon)^{2L}\,\,
e^{-\gamma_\epsilon \ln (x^2\, \Lambda^2 (\epsilon))
}\nonumber\\
&& =
(I_\epsilon)^{2L} \,\,{\rm exp}\,[\,\, \ln(1+ A_2
\lambda^2 + A_4 \lambda^4 +A_6 \lambda^6\cdots )\,\,] \ ,
\eea
where $I_\epsilon$ denote the Euclidean scalar
propagator in the position space
\be
I_\epsilon = \int {d^{2\omega} p\over (2\pi)^{2\omega}}
{1\over p^2} \, \, e^{i p \cdot x} =
{\Gamma(\omega-1)\over 4\pi^\omega}\,\, {1\over (x^2)^{\omega-1}}
\,.
\ee

Here we consider all Feynman diagrams, connected or not, contributing to a given order of $\lambda^{2 \ell}$. As the definition of the anomalous dimension matrix takes the logarithm, it suffices to compute connected diagrams only.
The anomalous dimension matrix is then extractable
as coefficient of $\ln (x^2)$ in the exponent.
Up to 6-loop orders, the spin chain Hamiltonians $H_{2 \ell}$ classified in section 2
are given by
\bea
&& H_2 = -\lim_{\epsilon\rightarrow 0}\,\,
\epsilon \,\, A_2\nonumber\\
&& H_4 =
-\lim_{\epsilon\rightarrow 0}\,\,
2\epsilon \,\, \Big[A_4- {1\over 2} A_2^2\,\,\Big]\,\nonumber\\
&& H_6 =
-\lim_{\epsilon\rightarrow 0}\,\,
3\epsilon \,\, \Big[\,\,A_6- {1\over 2} (A_2 A_4+A_4 A_2)
+ {1\over 3} A^3_2\,\,
\Big]\,.
\label{sixloop}
\eea
Note the extra factor $n$ multiplied. It arises combinatorially from
extracting coefficients of $\ln\, x^2$ from the $2 \ell $-loop contribution $\big(\ln\, A\,\big)_{2\ell}$,
\be
\big(\ln\, A\,\big)_{2\ell}=\left[\, c_{2\ell}\,\,\epsilon^{-1} +
O(\epsilon^0)\right] (x^2)^{\ell \,\epsilon}
=\left[\, c_{2\ell}\,\, {\epsilon^{-1}} + O\big(\epsilon^0\big)\,\right]
\left[ 1+ {\ell \,\epsilon}\,\, \ln x^2 + O(\epsilon^2)\right]\,.
\ee
At $2 \ell$-loop order, we have
\be
H_{2 \ell }= - \ell \,\,\lim_{\epsilon\rightarrow 0} \,\, \epsilon\,\,\big({\ln\, A}\big)_{2\ell}\, .
\ee
Summing over all loops,
\bea
H(\lambda^2) = {1 \over 2} + \sum_{\ell=1}^\infty \lambda^{2 \ell} \ H_{2 \ell} \qquad
\mbox{and} \qquad
A(\epsilon) = 1+ \sum_{\ell = 1}^\infty \lambda^{2 \ell} \ A_{2 \ell}(\epsilon)
\ee
and we have
\be
\int^1_0 {\rmd t\over t}\Big(H(\lambda^2\, t)-{1\over 2}\Big)
= -\lim_{\epsilon\rightarrow 0}\,\,
\epsilon  \,\,
\ln \, A(\epsilon)\, = - \lim_{\epsilon \rightarrow 0} \,\, 2 \epsilon \,\, \ln \, I_{\rm bare} (\epsilon).
\ee
Thus, starting from a different prescription, we have arrived at the same definition of the dilatation operator as (\ref{H-relation-CS}).
Here, the last equality followed from the fact that the $A(\epsilon)$ is defined in terms of the two-point correlation function $\langle\,
:\!\mathbb{O}\,(x):\, :\!\mathbb{O}\,(0):\,\rangle$,
so is square of the $I_{\rm bare} (\epsilon)$ (equivalently, the $\ln A(\epsilon)$ is twice of the $\ln I_{\rm bare}(\epsilon)$ connected diagrams) we considered in subsections 3.2 and 3.3.

Multiplicative renormalizability of the composite operators asserts that all $\epsilon^{-n}, (n \ge 2)$
singularities must cancel each other. We shall use such cancellations as a checkpoint of our computations.
In $A_2$, the contribution starts from $\epsilon^{-1}$ power and $H_2$ is automatically finite. In $A_4$, leading singularity starts from  $\epsilon^{-2}$ power. The coefficient of this singularity in $(\ln A)_4$
should vanish and the coefficient of order ${\epsilon^{-1}}$ leads to $H_4$.
In $A_6$, there are in general singularities of $\epsilon^{-3}$ and $\epsilon^{-2}$ powers.
The coefficients of them in
$(\ln A)_6$ vanish and the coefficient of order ${\epsilon}^{-1}$ leads to $H_6$.
As discussed in Section 2, beginning at 6-loop orders, several independent spin chain operators can
appear in $H$. The cancelation of higher-order singularities must then take place for the coefficients of
each independent operators. At 6-loops, this will provide an additional stringent consistency check of our earlier 4-loop results \cite{BakMinRey} on whether computation of $A_4$ was correct and identification of spin chain operators was complete.

\begin{figure}[ht!]
\centering \epsfysize=9cm
\includegraphics[scale=0.8]{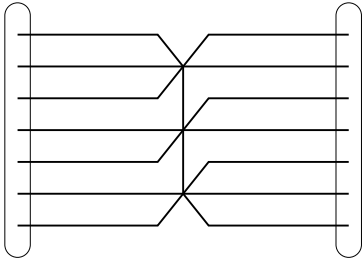}
\caption{\small \sl 6-loop homogeneous diagram involving maximal
shuffling terms}
\label{fig1}
\end{figure}

\begin{figure}[ht!]
\centering \epsfysize=9cm
\includegraphics[scale=0.8]{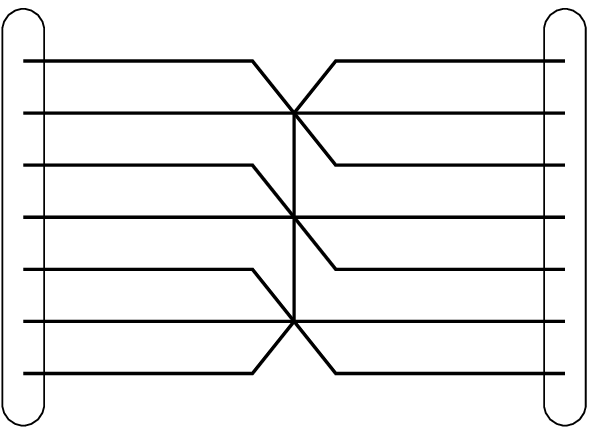}
\caption{\small \sl
adjoint of the above homogeneous diagram}
\label{fig2}
\end{figure}

\begin{figure}[ht!]
\centering \epsfysize=9cm
\includegraphics[scale=0.8]{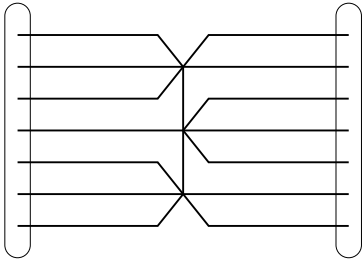}
\caption{\small \sl
6 loop inhomogeneous diagram whose interactions are maximal-ranged  }
\label{fig3}
\end{figure}

\begin{figure}[ht!]
\centering \epsfysize=9cm
\includegraphics[scale=0.8]{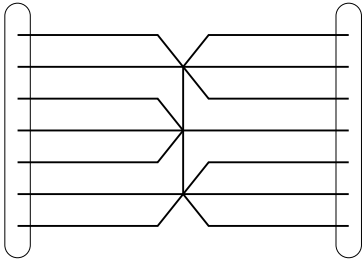}
\caption{\small \sl
adjoint of the above inhomogeneous diagram }
\label{fig4}
\end{figure}

\section{Operator Contents of 6-Loop Dilatation Operator}
We asserted earlier that, beginning at 6-loop order, the dilatation operator
becomes complicated because there arise two types of maximally ranged interactions.
In this section, we shall explain this in detail by analyzing operator contents of
the 6-loop dilatation operator, focusing on the maximally ranged diagrams involving
seven lattice sites.

The first type of spin chain operator arises from the homogeneous diagrams and they
give rise to the maximal shuffling terms. The diagrams in Figs.~\ref{fig1} and
\ref{fig2} belong to this category. The sextet interaction vertices are combined as
in  Figs.~\ref{fig1} and \ref{fig2}. The second type arises from
the inhomogeneous diagrams. They are still maximally ranged but do not generate
the maximal shuffling terms. Figs.~\ref{fig3} and
\ref{fig4} depict the inhomogeneous maximal ranged diagram
at 6-loop order.

Their operator contents 
 can be constructed from the basic building structure
arising from the 2-loop sextet scalar vertices:
\bea
\mathbb{O}^{\ell}_{{123}}&=& 2\mathbb{I} -
\mathbb{K}_{\ell+1,\, \ell+2}
-\mathbb{K}_{\ell+2,\,\ell+3} +2 \mathbb{P}_{\ell+1,\, \ell+3}
\mathbb{K}_{\ell+1,\, \ell+2}\nonumber\\
&+& 2 \mathbb{P}_{\ell+1,\, \ell+3} \mathbb{K}_{\ell+2,\, \ell+3}- 4
 \mathbb{P}_{\ell+1,\, \ell+3}\,,
\eea
where $\mathbb{K}_{ab}^{a'b'}= \delta_{ab}\,\delta^{a'b'}$ is the
${\bf 4}\!-\!\overline{\bf 4}$ covariant contraction operator.
Here $\ell$ could be even or odd but, as mentioned before,
 we focus on  the odd chain where
$\ell$ is even.
Acting this operator on the set of states in (\ref{seto}),
all the terms involving the contraction operators
drop out and the operator is reduced to
\be
\mathbb{O}^{\ell}_{{123}}=2\mathbb{I} - 4
 \mathbb{P}_{\ell+1,\, \ell+3}\,.
\ee
Below we shall present this reduced part  only
and omit all the contraction type operators.
The operator structure of the diagram in Fig.~\ref{fig1} is then
\be
\mathbb{O}^{\ell}_{{123}}\,\,\mathbb{O}^{\ell}_{345}\,\,
 \mathbb{O}^{\ell}_{567}
=
2^3\, (\mathbb{I}-2 \mathbb{P}_{\ell+1,\, \ell+3})\,
(\mathbb{I}-2 \mathbb{P}_{\ell+3,\, \ell+5})\,
(\mathbb{I}-2 \mathbb{P}_{\ell+5,\, \ell+7})\, .
\ee
Similarly,  the operator structure of Fig.~\ref{fig2} 
 becomes
\be
\mathbb{O}^{\ell}_{{567}}\,\,\mathbb{O}^{\ell}_{345}\,\,
 \mathbb{O}^{\ell}_{123}
=
2^3\, (\mathbb{I}-2 \mathbb{P}_{\ell+5,\, \ell+7})\,
(\mathbb{I}-2 \mathbb{P}_{\ell+3,\, \ell+5})\,
(\mathbb{I}-2 \mathbb{P}_{\ell+1,\, \ell+3})\,.
\ee
The numerical parts of  these Feynman diagrams
are the same. So, adding the two contributions, we obtain the spin chain operator
\be
\mathbb{O}^{\ell}_{1\sim 7}=
\mathbb{O}^{\ell}_{123}\,\mathbb{O}^{\ell}_{345}\,
\mathbb{O}^{\ell}_{567} +
\mathbb{O}^{\ell}_{567}\,\mathbb{O}^{\ell}_{345}\,
\mathbb{O}^{\ell}_{123}\,.
\ee
One can show that
this operator is related to
the $\mathbb{O}^{\ell}_{6,6}$ by
\be
\sum_{\ell}\,\,\mathbb{O}^{\ell}_{1\sim 7}=
64\,\sum_{\ell} \,\Big[
\, 64 \mathbb{O}^{\ell}_{6,6} -80 \mathbb{O}^{\ell}_{4,4}
+22 \mathbb{O}^{\ell}_{2,2} - {1\over 4}\, \mathbb{I}\,
\Big]\,.
\label{identity66}
\ee
The details of the derivation is relegated to appendix B.
We focus on the maximally ranged interaction part only, and omit
interactions of lower range given by 
$e_{6,2}\, \mathbb{O}_{2,2}$ and $e_{6,4}\,\mathbb{O}_{4,4}$.
Thus, we conclude that the operator content of
the homogeneous diagram is given by
\be
\mathbb{O}^{\ell}_{1\sim 7}=  4^6\,\, \mathbb{O}^{\ell}_{6,6}
 + \cdots
\label{sixty4}
\ee
where the ellipses denote the interactions of lower range.

The operator structure of the diagram in Fig.~\ref{fig3} 
 is identified as
\be
\mathbb{O}^{\ell}_{{567}}\,\,\mathbb{O}^{\ell}_{123}\,\,
 \mathbb{O}^{\ell}_{345}
=
2^3\, (\mathbb{I}-2 \mathbb{P}_{\ell+5,\, \ell+7})\,
(\mathbb{I}-2 \mathbb{P}_{\ell+1,\, \ell+3})\,
(\mathbb{I}-2 \mathbb{P}_{\ell+3,\, \ell+5})\,,
\ee
and that of the diagram in Fig.~\ref{fig4} 
as
\be
\mathbb{O}^{\ell}_{{345}}\,\,\mathbb{O}^{\ell}_{123}\,\,
 \mathbb{O}^{\ell}_{567}
=
2^3\, (\mathbb{I}-2 \mathbb{P}_{\ell+3,\, \ell+5})\,
(\mathbb{I}-2 \mathbb{P}_{\ell+1,\, \ell+3})\,
(\mathbb{I}-2 \mathbb{P}_{\ell+5,\, \ell+7})\, .
\ee
Notice that the order the basic unit operators are multiplied
is different from that in the homogeneous diagrams.
Again, the numerical parts of these Feynman diagrams
are the same. Combining them together, we obtain
\be
\tilde{\mathbb{O}}^{\ell}_{1\sim 7}=
\mathbb{O}^{\ell}_{567}\,\mathbb{O}^{\ell}_{123}\,
\mathbb{O}^{\ell}_{345} +
\mathbb{O}^{\ell}_{345}\,\mathbb{O}^{\ell}_{123}\,
\mathbb{O}^{\ell}_{567}\,.
\ee
In appendix B, we show that
\bea
\widetilde{\mathbb{O}}^{\ell}_{1\sim 7}
&=& - \mathbb{O}^{\ell}_{1\sim 7}
-\, 64\, \mathbb{P}_{\ell+1,\, \ell+7}+ 32\, \mathbb{I} \nonumber\\
&=& - \mathbb{O}^{\ell}_{1\sim 7}+\,64\,\Big[\,\,
64\, \widetilde{\mathbb{O}}^{\ell}_{6,6} -96 \mathbb{O}^{\ell}_{4,4}
+36 \mathbb{O}^{\ell}_{2,2} - {1\over 2}\, \mathbb{I}\,\,\Big]\,.
\label{identityt66}
\eea
Hence, we have
\be
\widetilde{\mathbb{O}}^{\ell}_{1\sim 7}=
 4^6\, \big[-\mathbb{O}^{\ell}_{6,6}+
\tilde{\mathbb{O}}^{\ell}_{6,6}
\,\big]+ \cdots
\label{sixty44}
\ee
where ellipses denotes the interactions of shorter range given by
$e_{6,2}\, \mathbb{O}_{2,2}$ and $e_{6,4}\,\mathbb{O}_{4,4}$.

\begin{figure}[ht!]
\centering \epsfysize=9cm
\includegraphics[scale=0.8]{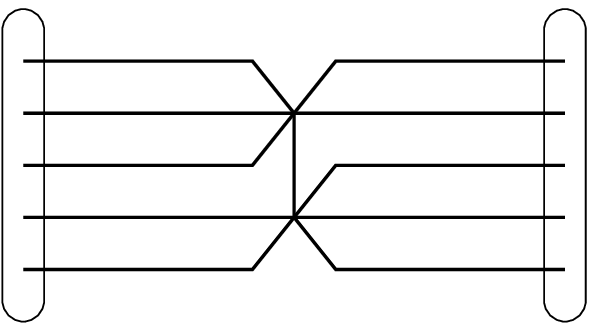}
\caption{\small \sl
4-loop homogeneous diagram involving  maximal shuffling terms}
\label{fig5}
\end{figure}

\begin{figure}[ht!]
\centering \epsfysize=9cm
\includegraphics[scale=0.8]{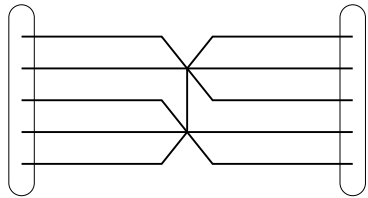}
\caption{\small \sl
adjoint of the above 4-loop homogeneous diagram}
\label{fig6}
\end{figure}

For completeness, we review the operator structures arising in the 4-loop maximally
ranged interactions. For the 4-loop order, only diagrams of homogeneous type are present.
These Feynman diagrams are depicted in Fig.~\ref{fig5} and Fig.~\ref{fig6}.
The numerical parts of the 4-loop amplitudes
in Figs.~\ref{fig5} and \ref{fig6} 
are the same. Combining them, we have the spin chain operator
\bea
&& \mathbb{O}^{\ell}_{1\sim 5}= \mathbb{O}^{\ell}_{123}\, \mathbb{O}^{\ell}_{345}+
\mathbb{O}^{\ell}_{345} \, \mathbb{O}^{\ell}_{123}
\nonumber\\
&& =16\,\Big[\, {1\over 2} \mathbb{I} -\mathbb{P}_{\ell+1,\, \ell+3} -
\mathbb{P}_{\ell+3,\, \ell+5} +
\mathbb{P}_{\ell+1,\, \ell+3}\, \mathbb{P}_{\ell+3,\, \ell+5}+
\mathbb{P}_{\ell+3,\, \ell+5}\,\mathbb{P}_{\ell+1,\, \ell+3}\, \Big]\,.
\eea
This
can be rewritten as
\be
\mathbb{O}^{\ell}_{1\sim 5}= 16 \mathbb{P}_{\ell+1,\, \ell+5} - 8 \mathbb{I} \ ,
\label{identity55}
\ee
where we again omit the contraction type operators. Here, we used the identity
\be
\epsilon_{I\,a_1 a_3 a_5}\,\,\epsilon_{I\, b_1b_3 b_5} =  \mathbb{I}
 +
\mathbb{P}_{13}\,\,\mathbb{P}_{35}
+
\mathbb{P}_{35}\,\,\mathbb{P}_{13}
- \mathbb{P}_{13}-\mathbb{P}_{35} -\mathbb{P}_{15}=0\,,
\label{identity}
\ee
taking account of the fact that $a_\ell$ and $b_\ell$ are running over only $\{1,2\}$.

\section{Computation in the 6-Loop Dilatation Operator}
In this section, we shall compute maximal-ranged interactions in the 6-loop dilatation operator
(\ref{sixloop}) and extract the coefficients  $e_{6,6}$ and $\kappa_6$.
As explained in the previous section, it suffices to compute the 6-loop
maximally ranged operators,
$\mathbb{O}_{6,6}$ and $\tilde{\mathbb{O}}_{6,6}\,\,$.

\begin{figure}[ht!]
\centering \epsfysize=9cm
\includegraphics[scale=0.8]{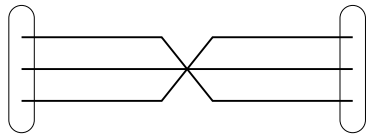}
\caption{\small \sl
2-loop diagram involving three sites
}
\label{fig7}
\end{figure}

The first contribution to the maximal-range interactions
arise from 
the $A_2^3$ part.
The relevant 2-loop amplitude in Fig.~\ref{fig7}
is organized as follows:
\be
A_2 = a_2(\epsilon) \,\, (x^2 \pi)^{3-2\omega}\,\, \Big[\,\,\cdots +
\mathbb{O}^{\ell}_{123}+
\mathbb{O}^{\ell}_{345}+\mathbb{O}^{\ell}_{567} + \cdots\,\,
\Big]\,.
\label{A2}
\ee
In Ref.~\cite{BakMinRey}, the numerical  coefficient $a_2(\epsilon)$ was obtained as
\bea
&& a_2(\epsilon)  =
-{\big[\,\Gamma(\omega -1)\,\big]^3 \big[\Gamma(3-2\omega)\big]^2
\Gamma(5\omega -6)
\over 16\pi \,\,\,\big[\Gamma(3\omega-3)\big]^2 \,\,\Gamma(6-4\omega)
}\nonumber\\
&&   \,
 = -{1\over 4\epsilon}\left[1+ \Bigl(1-\psi(
\mbox{\small ${1\over2}$}
)
\Bigr)\epsilon
+ {1\over 24} \Bigl(11\pi^2 + 12\psi^2(
\mbox{\small ${1\over2}$}
)
-24 \psi(
\mbox{\small ${1\over2}$}
)
-72
\Bigr)\epsilon^2+
O(\epsilon^3)\right]
\,,
\eea
where the poly-gamma functions take the value $\psi(\mbox{\small
${1\over 2}$})=-{\bf C}-2\ln 2$ and
$\psi(1)=-{\bf C}$ with ${\bf C}\,\,(=0.577215\cdots)$ being the Euler's constant.
From this, we extract contribution of the $A_2^3$ to the relevant maximal-ranged interactions
as
\be
A^3_2 =
\big(\, a_2(\epsilon)\,\big)^3 \,\, (x^2 \pi)^{9-6\omega}\,\, \Big[\,\,\cdots +
\mathbb{O}^{\ell}_{1\sim 7}+
 2\,\tilde{\mathbb{O}}^{\ell}_{1\sim 7} + \cdots\,\,
\Big]\,,
\ee
where we already omit all disconnected contributions.

The $(\,A_2\, A_4+ A_4\, A_2\,)$ part is another contribution to
the maximal ranged interactions. The relevant 4-loop contribution to
the maximal-ranged interactions has the structure
\be
A_4 =  (x^2 \pi)^{6-4\omega} \Big[ \cdots + a_4(\epsilon)\,\,\big[
\mathbb{O}^{\ell}_{1\sim 5}+
  \tilde{\mathbb{O}}^{\ell}_{3\sim 7}\,\big]+ a^2_2(\epsilon)
\,\,\mathbb{O}^{\ell}_{123}\,\,
\mathbb{O}^{\ell}_{567} + \cdots
\Big]\,.
\ee
This along with (\ref{A2}) gives rise to six loop maximal-ranged interactions  through
$(\,A_2\, A_4+ A_4\, A_2\,)$. We find
\be
A_2\, A_4+ A_4\, A_2=
(x^2 \pi)^{9-6\omega} \Big[ \cdots + 2\,a_2(\epsilon)
\,a_4(\epsilon)\,\,\big[\,\,\mathbb{O}^{\ell}_{1\sim7}+
\tilde{\mathbb{O}}^{\ell}_{1\sim 7}\,\big]  +
a^3_2(\epsilon)\,
\,\tilde{\mathbb{O}}^{\ell}_{1\sim 7} + \cdots
\Big]\,.
\ee

Lastly, as discussed before, the relevant maximal-range interactions in $A_6$ are
given by
\be
A_6 =
(x^2 \pi)^{9-6\omega} \Big[ \cdots + a_6(\epsilon)\,\,\mathbb{O}^{\ell}_{1\sim7}+
  \tilde{a}_6(\epsilon)\,\,\tilde{\mathbb{O}}^{\ell}_{1\sim 7} + \cdots
\Big]\,.
\label{6loopa6}
\ee
Using (\ref{sixloop}), we found that the coefficients of the
maximal-ranged operators in $(\ln A)_6$ has the structure
\bea
 (\ln A)_6 &=& \Big[a_6(\epsilon) -a_2(\epsilon) a_4(\epsilon) +
{1\over 3} a_2^3(\epsilon)
\Big]\,\,
\mathbb{O}_{1\sim 7}\nonumber\\
&+&\Big[\tilde{a}_6(\epsilon) -a_2(\epsilon) a_4(\epsilon) +{1\over
 6} a_2^3(\epsilon)
\Big]\,\,
\tilde{\mathbb{O}}_{1\sim 7}+\cdots\,.
\label{loga}
\eea

\begin{figure}[ht!]
\vspace{0.5cm}
\centering \epsfysize=10cm
\includegraphics[scale=0.7]{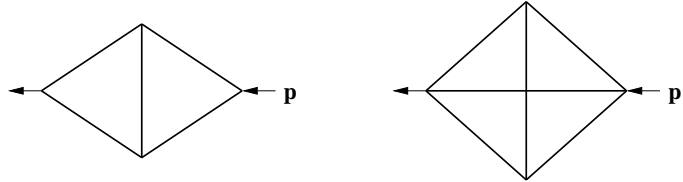}
\caption{\small \sl
The left is a skeleton 2-loop diagram used to evaluate $A_4$. The
skeleton 4-loop diagram on the right is for $A_6$. We inject a momentum
$p$ following the arrows in each diagram.
}
\label{skeleton}
\end{figure}

The 4-loop coefficient $a_4(\epsilon)$ 
was determined in~\cite{BakMinRey}
 \be
a_4(\epsilon)= J_1(\epsilon)\cdot
{\big[\,\Gamma(\omega-1)\big]^5\over 4^4 \pi^2} \left[
{\Gamma(2-\omega) \Gamma(3-2\omega)\over \Gamma(2\omega-2)
\Gamma(3\omega-3)}
\right]^2 \cdot {\Gamma(9\omega -11)\over
\Gamma(11-8\omega)}\,.
\ee
The integral $J_1(\epsilon)$ is introduced
by the following skeleton 2-loop
integral ${K}_5$:
\bea
{K}_5(w_1, w_2, w_3, w_4, w_5)&=& (4\pi)^{2\omega}\,
(p^2)^{-2\omega +
\sum^5_{k=1}w_k}
\int {d^{2\omega} k\over (2\pi)^{2\omega}}
{d^{2\omega} l
\over (2\pi)^{2\omega}}
{ 1
\over (k^2)^{w_1} (l^2)^{w_2}
} \nonumber\\
&&\ \ \ \otimes {1\over \big[(l-p)^2\big]^{w_3}
\big[(k-p)^2\big]^{w_4} \big[(k-l)^2\big]^{w_5}} \,.
\label{k5}
\eea
The corresponding skeleton 2-loop integral is depicted in Fig.~\ref{skeleton}.
As we evaluate the two-point correlation functions in the $x$
space, we first evaluate the skeleton graph with nonvanishing momentum injected
from the left to the right and then Fourier transform it back to the
$x$-space.  See Ref.~\cite{BakMinRey}  for the details of the method.
The integral $J_1(\epsilon)$ evaluated from the skeleton 2-loop integral $K_5$:
\bea
 J_1(\epsilon) \equiv K_5(2-\omega,3-2\omega, 2-\omega,
3-2\omega,\,\,1\,\,) \ .
\label{twoloop-1}
\eea
This integral is already nontrivial and does not allow analytic evaluation. Here we
 evaluated it numerically by using a Mathematica package. We took the strategy
of expanding the integral
$J_1(\epsilon)$ in powers of $\epsilon$ and determine term by term. The result is
\bea
&& J_1(\epsilon) \equiv
{\alpha^{-1}_{J_1}\over \epsilon}\Big[
1+ \alpha^{0}_{J_1}\epsilon + \alpha^{1}_{J_1}\epsilon^2 +
  \cdots \Big]
\nonumber\\
 &&\  =-{2\over 3\pi \,\, \epsilon}\Big[ 1+ \Big(
13+ \psi(
\mbox{\small ${1\over2}$})
\Big) \epsilon +\Big( 488-{17\over 2}\pi^2 + 52 \psi(
\mbox{\small ${1\over2}$})+2\psi^2(
\mbox{\small ${1\over2}$}) \Big) {\epsilon^2 \over 4} + \cdots
\Big]\,.
\label{resultJ1}
\eea
We computed the leading coefficient $\alpha^{-1}_{J_1}$ numerically
and matched to the above rational value. This is the value required for the
multiplicative renormalizability of composite operator at four loops.
The second coefficient is the one relevant for the 4-loop
amplitude $\alpha^{0}_{J_1}$. Here, we checked  this
against the requirement of the multiplicative renormalizability at six loops. We computed the last coefficient $\alpha^{1}_{J_1}$ numerically up to 30 significant digits. We  match this number as a linear
combination of $1$ and transcendental numbers $\pi^2$, $\psi(
\mbox{\small ${1\over2}$})$, $\psi^2(
\mbox{\small ${1\over2}$})$ with the requirement that the coefficients are rational numbers.
This is highly nontrivial to check. However, utilizing the remarkable algorithm PSLQ discovered recently
by Ferguson, Baily and Arno, we determined the combination in (\ref{resultJ1}) as the unique solution. The details are explained in appendix D.
Using this, we found that  $a_4(\epsilon)$ has the series expansion
\be
 a_4(\epsilon) = {1\over 32 \epsilon^2}\,\, \Big[\, 1+
\Big(\,\, 3 -2\,\psi(
\mbox{\small ${1\over2}$})\,\Big)\,\,  \epsilon 
+ \Big(\,\,
 -2+{11\over 12}\,\pi^2
+2\,\psi^2(
\mbox{\small ${1\over2}$})-6\,\psi\,(
\mbox{\small ${1\over2}$})\,\,\Big)\,\,
\epsilon^2+\cdots
\Big]\,.
\ee

We now turn to the 6-loop coefficients $a_6(\epsilon)$ and $\tilde{a}_6(\epsilon)$.
Consider first the coefficients arising from the diagrams in Figs.~\ref{fig1} and \ref{fig2}.
One can easily verify that these two coefficients
are the same.
As explained in appendix C,
we find
\bea
 a_6(\epsilon) = 
-{J_4(\epsilon)\over
4^6\pi^3}\,\,\,
{\big[\,\Gamma(\omega -1)\,\big]^7  \,\,\big[\,\Gamma(2-\omega)\,\big]^4\,\,
\big[\,\Gamma(3-2\omega)\,\big]^2 \,\,\Gamma(13\omega -16)
\over
\big[\,\Gamma(2\omega-2)\,\big]^4\,\,
\big[\,\Gamma(3\omega-3)\, \big]^2 \,\,\Gamma(16-12\omega)
}
\,.
\label{a6}
\eea
The Feynman integral can be reduced analytically to the function $J_4(\epsilon)$. Consider the skeleton 4-loop integral defined by
\bea
 K_8(w_1 \cdots  w_8) &\equiv &
(4\pi)^{4\omega}\,\,
(p^2)^{-4\omega +
\sum^8_{k=1}w_k}\,
 \int {d^{2\omega} k_1 \over (2\pi)^{2\omega}}
{d^{2\omega}k_2
\over (2\pi)^{2\omega}}
{d^{2\omega} k_3 \over (2\pi)^{2\omega}}
{d^{2\omega} k_4 \over (2\pi)^{2\omega}}
\nonumber\\
&\otimes &
{
1
\over (k_1^2)^{w_1}\,\, (k_2^2)^{w_2}\,\,
(k_3^2)^{w_3}\,\, (k_4^2)^{w_4}\,\,
\big[\,(k_1 + k_2-p)^2\big]^{w_5} }
\nonumber\\
& \otimes &
{1\over
\big[\,(k_3 + k_4-p)^2\big]^{w_6}\,\,
\big[\,(k_1-k_3)^2\big]^{w_7} \,\,
\big[\,(k_2-k_4)^2\big]^{w_8}
}
\label{qloop-2}
\eea
Then the function $J_4(\epsilon)$ is given by
\bea
 J_4(\epsilon) = K_8(2-\omega,3-2\omega,3-2\omega, 2-\omega,
 2-\omega, 2-\omega, \,\,1\,\ , \,\,1\,\,)
\label{sixj4}\,.
\eea
By using the PSLQ algorithm again, we determined that
the function $J_4(\epsilon)$ has the expansion in powers of $\epsilon$ as
\be
 J_4(\epsilon) &=& {16\over 135 \pi^2 \epsilon^2}\,\, \Big[\, 1+
\Big(\,\, 28+ {14\over 15} +2\psi(
\mbox{\small ${1\over2}$})\,\Big)\,\,  \epsilon \nonumber\\
&+& \Big(\,\,522-{11\over 25}- {2\over 9}
-{27\over 2}\,\pi^2 +
{868\over 15}\,\psi(
\mbox{\small ${1\over2}$})
+ 2\,\psi^2\,(
\mbox{\small ${1\over2}$})\Big)\,\,
\epsilon^2\,\,\,
\Big]\,.
\label{j4k}
\ee
The coefficients of the first two terms fit precisely to the values
required for the multiplicative renormalizability. As presented in appendix D, we checked numerically the first coefficient up to 14 significant digits and the second up to 9 significant digits.
From the expansion (\ref{j4k}), we found that $a_6(\epsilon)$ is given by
\be
 a_6(\epsilon) = -{1\over 384 \epsilon^3}\,\, \Big[\, 1+
3 \cdot \Big(\,\, 2 -\psi(
\mbox{\small ${1\over2}$})\,\Big)\,\,  \epsilon 
+ \Big(\,\,{80\over 9}
+{11\over 9}\,\pi^2 +
4\,\psi^2(
\mbox{\small ${1\over2}$})-16\,\psi\,(
\mbox{\small ${1\over2}$})\,\,\Big)\,\,
\epsilon^2\,\,\,
\Big]\,.
\ee
Adding all pertinent contributions to $\mathbb{O}^\ell_{6,6}$, we get
\be
a_6(\epsilon) -a_2(\epsilon) a_4(\epsilon) +{1\over 3} a_2^3(\epsilon)
= -{1\over 96\epsilon} + O(\epsilon^0)\,.
\ee
Contributions to the $\epsilon^{-3}$ and
$\epsilon^{-2}$ terms cancel out, thus satisying the consistency of multiplicative
renormalizability. From (\ref{sixloop}) and $(\ref{sixty4})$, we finally determined
the coefficient of $\mathbb{O}_{6,6}^\ell$ as
\be
e_{6,6}= 128\, .
\ee
This is the value that agrees exactly with the prediction of the integrability.

Similarly, the coefficient $\tilde{a}_6(\epsilon)$
in (\ref{6loopa6}) arises from  the diagrams in Figs.~\ref{fig3}
and  \ref{fig4}. As explained in appendix C,
we find
\bea
 \tilde{a}_6(\epsilon) = 
-{J_5(\epsilon)\over
4^6\pi^3}\,\, {\big[\,\Gamma(\omega -1)\,\big]^6 \,\,
\big[\,\Gamma(2-\omega)\,\big]^2\,\,
\big[\,\Gamma(3-2\omega)\,\big]^3\,\, \Gamma(13\omega -16)
\over
\big[\,\Gamma(2\omega-2)\,\big]^2\,\,
\big[\,\Gamma(3\omega-3))\,\big]^3\,\, \Gamma(16-12\omega)
}
\,,
\label{ta6}
\eea
where the function $J_5(\epsilon)$ is the skeleton 4-loop integral:
\bea
  J_5(\epsilon) = K_8(2-\omega,2-\omega,3-2\omega, 3-2\omega,
 3-2\omega, \,\, 1\,\,, \,\,1\,\ , \,\,1\,\,)\,.
\label{j5k}
\eea
Again, utilizing the PSLQ algorithm, we computed the series expansion of  $J_5(\epsilon)$ as
\be
J_5(\epsilon) &=& {16\over 135  \epsilon}\,\, \Big[ 1+
\Big(19+  {14\over 15}+{9\over 2}
+\psi(1)+\psi(
\mbox{\small ${1\over2}$})\Big)\, \epsilon \nonumber\\
&+& \Big(\,
{177587\over 225}
-{61\over 6}\,\,\pi^2 +
{733\over 15}\,\,\big[\,\psi(1)+
\psi(
\mbox{\small ${1\over2}$})\big]+
\big[\, \psi(1)+\psi(
\mbox{\small ${1\over2}$})\,\big]^2\,\, \Big)
\,\,{\epsilon^2 \over 2} +\cdots \Big]\,.
\ee
We confirmed that the first two coefficients are the values required by the multiplicative
renormalizability. See appendix D.
We then obtained the coefficient $\tilde{a}_6(\epsilon)$ in $\epsilon$ expansion as
\be
 \tilde{a}_6(\epsilon) = -{1\over 192 \epsilon^3}\,\, \Big[\, 1+
\Big(\,\, {9\over 2} -3\psi(
\mbox{\small ${1\over2}$})\,\Big)\,\,  \epsilon 
+ \Big(\,\,
{11\over 8}\,\pi^2 +{9 \over 2} \,\psi^2(
\mbox{\small ${1\over2}$})-{27 \over 2} \,\psi\,(
\mbox{\small ${1\over2}$})\,\,\Big)\,\,
\epsilon^2\,+
\cdots \Big]\,.
\ee
From this, we find the pertinent contribution to $\widetilde{O}_{6,6}^\ell$ as
\be
\tilde{a}_6(\epsilon) -a_2(\epsilon) a_4(\epsilon) +{1\over 6} a_2^3(\epsilon)
= 0\, \cdot\, {1\over\epsilon} + O(\epsilon^0)\,.
\ee
Again the coefficients of $\epsilon^{-3}$ and
$\epsilon^{-2}$ are vanishing, which is the requirement of the
renormalizability. From (\ref{sixloop}) and $(\ref{sixty44})$, we finally find that
\be
\kappa_{6}= 0\,.
\label{kappa6}
\ee
As such, we conclude the operator $\widetilde{\mathbb{O}}_{6,6}$, even though the integrability and the excitation symmetry allow it to be present, does not contribute to the 6-loop dilatation operator.
Curiously, this parallels exactly to the structure of the dilatation operator in ${\cal N}=4$
super Yang-Mills theory~\cite{Serban:2004jf} --- in both theories, the maximally-ranged interactions are governed by a single spin chain operator.

Summarizing, we determined uniquely the operator contents and the recursive structure of the 6-loop dilatation operator given by
\be
H_6= 128\,\,\sum_{\ell}\mathbb{O}^\ell_{6,6} + \cdots \ ,
\ee
where the ellipses denotes shorter-ranged terms we omitted.

\section{Remarks on ABJ Theory}
Up to this point, we focused, among the ${\cal N}=6$ Chern-Simons theory, on the ABJM theory. This theory has the gauge group $U(N) \times U(N)$, the Chern-Simons coefficients $+k, -k$ and is invariant under parity. Notice that the parity is generalized to interchange the two gauge groups and hence all matter fields and their conjugates.

We can extend straightforwardly our considerations to the ABJ theory, which has the gauge group $U(M) \times U(N)$, the Chern-Simons coefficients $+k, -k$ but with $N < M < N+k$. This theory then breaks the generalized parity. As explained in \cite{Bak2} and recapitulated in section 2, the integrability and the Yang-Baxter equations therein implied that the dilatation operator is in general parity non-conserving. Nevertheless, explicit computations in \cite{Bak2, BakMinRey} showed that the dilation operator sustains to be parity conserving. At 6-loop order and beyond, as there begins to arise two types of spin chain operators in the dilatation operator, it becomes interesting if they source a room for parity (non)conservation beginning at this order.

From the Feynman diagrammatics, however, it is evident that all the changes in 6-loop diagrams of the ABJ theory compared to those in the ABJM theory is rather trivial; one just replaces the 't Hooft coupling-squared $\lambda^2$ of the ABJM theory by the product of two 't Hooft couplings $\lambda \overline{\lambda}$ of the ABJ theory. The nullification result $\kappa_6 = 0$ in (\ref{kappa6}) of the
ABJM theory was {\sl not} sensitive to the `t Hooft coupling as its dependence is an overall weighting factor. In the ABJ theory, by repeating the color factor counting in the Feynman diagrams, we find that
all the contributing diagrams have the common dependence on $M$ and $N$. Therefore, $\kappa_6 = 0$ in the
ABJ theory as well, viz. there arises a unique spin chain operator to the six-loop dilatation operator.
This then eliminates the possibility that, within maximal ranged interactions, parity non-conserving effect
arises at six loops and beyond.

On the other hand, we suspect that the effect
of parity non-conservation may show up at planar limit in some
of the higher conserved charges
${\cal Q}_n, \overline{\cal Q}_n \ (n \ge 3)$. With indications from recent results of \cite{nonplanar}, we also expect that parity non-conserving effects will also arise from non-planar corrections.
Given that there is no known examples of parity non-conserving yet integrable system, further understanding on this issue would be very rewarding.
\section{Conclusions}

In this paper, we studied the dilatation operator of ${\cal N}=6$ Chern-Simons theory,
paying attention to two pertinent issues that begins to arise from six loops and beyond:
magnon spectrum and operator contents of the dilatation operator viewed as a spin chain Hamiltonian.
The integrability together with the excitation symmetry plus the 2-loop results
led to the unique prediction 
for the maximal shuffling terms to all orders.

We computed the maximal shuffling terms in the dilatation operator. We found that the
coefficients of this term, which arise from so-called {\sl homogeneous} diagrams, agree with the
prediction of the integrability and the recursive property. We thus found that the situation of maximal shuffling interactions is exactly parallel to the situation in ${\cal N}=4$ super-Yang-Mills theory.
Despite the parallel, we also argued that the ${\cal N}=6$ Chern-Simons theory is significantly different
from the ${\cal N}=4$ super Yang-Mills theory since there separately exist so-called {\sl inhomogeneous}
diagrams. These diagrams do not generate maximal shuffling terms but are still maximally-ranged. We showed
that the {\sl inhomogeneous} diagrams are not recursive and appears to depend on the choice of infrared regularization.

To handle these difficulties, we adopted {\sl ab initio} method of extracting the dilatation operator from
two-point correlation functions of single-trace composite operator. In dual Type IIA string theory, this amounts to deriving time-evolution Hamiltonian of a single free string. The method was particularly convenient for extracting
not only the spectrum but also the operator contents of the dilatation operator. By explicit computations, we determined the maximally-ranged interactions of the dilatation operator up to 6-loop order. We were able to compute pertinent Feynman diagrams by utilizing the remarkable integer-relation algorithm PSLQ. We found that, though details of Feynman diagrams contributing to the dilatation operator are very different, the operator contents and the recursive coefficients of the maximal-ranged interaction is identical to the
dilatation operator in the ${\cal N}=4$ super-Yang-Mills theory. This result is of course
consistent with the integrability, which  we consider as a highly nontrivial test for the ${\cal N}=6$ ABJM
theory.

We pointed out that extension of the results to the parity non-conserving ${\cal N}=6$ ABJ theory is rather trivial, at least for the maximally-ranged interactions. One just replaces the 't Hooft coupling
squared $\lambda^2$ of the ${\cal N}=6$ ABJM theory  by product of the two 't Hooft couplings $\lambda \overline{\lambda}$ of the ${\cal N}=6$ ABJ theory. The resulting dilatation operator remains to be parity
conserving.

In the above test, we assume  that the pseudo-momentum coincides with the lattice momentum in the operator space. This assumption can be further tested computing the next-maximal shuffling coefficient $e_{6,4}$, which is predicted as $e_{6,4}= -32 h_2$.

Recently, the coefficient $h_2$ was computed in \cite{Mina2}. Given that Ward identity was not verified to the order the computation is based on, the result of \cite{Mina2} needs to be checked independently. In particular, there is no first-principle understanding of its origin. One promising approach of the independent check would be the following. In  Ref.~\cite{BakMinRey}, it was noted that $h(\lambda)$ should
also be present as a renormalization factor in the central charges of the off-shell  $[psu(2|2)\oplus psu(2|2)]\ltimes \mathbb{R}^{2,1}$ superalgebra. Any further understanding of this function
$h(\lambda)$ such as renormalization of the central charges would be extremely interesting.

\section*{Acknowledgement}
We thank Gleb Arutyunov, Jin-Beom Bae, Niklas Beisert, Sergey Frolov, Carlo Meghenelli, Matthias Staudacher and Takao Suyama for helpful discussions. Part of results in this work was reported at {\sl Integrability in Gauge and String Theory} conference at Potsdam, Germany (June 29 - July 3, 2009).
This work was supported in part by the National Research Foundation of Korea Grants SRC-CQUEST-R11-2005-021, R01-2008-000-10656-0, 2005-084-C00003, 2008-313-C00175, 2009-008-0372 and U.S. Department of Energy Grant DE-FG02-90ER40542.

\appendix

\section{Comparative calculation of inhomogeneous diagram}
Below, for comparison with two-point function method, we shall illustrate a comparative calculation of a sample inhomogeneous diagram in operator-mixing method. Consider the inhomogeneous diagram in Fig.~3.
The diagram has three skeleton vertices but, by reflection symmetry of the diagram, there are only two independent skeleton vertices. To control potential infrared divergences,
it is necessary to inject nonzero momentum to the operator judiciously diagram by diagram. Here, 
consider injecting $q$ at the top or bottom skeleton vertex. 
This is what we did for the homogeneous diagrams and found to yield an infrared finite result.
For the skeleton 2-loop, which is actually elementary 6-loop, we find the contribution
\bea
{\cal B}_{2}^{\rm bare} &=& (- (2 \pi)^2)^3 (-4)^3 C_3^2 G(1 + \epsilon, 1)
G(3 - \omega + \epsilon, \epsilon) {1 \over (4 \pi)^{2 \omega}}
{1 \over (q^2)^{3 \epsilon}} \nonumber \\
&=&- {1 \over (4 \pi)^3}\,\, \Big[ {4 \pi \over q^2}
\Big]^{3 \epsilon}\,\,
{1\over 3!\,\, \epsilon^3}\,\,
{\big[\,\,\Gamma \big({1\over 2}- {\epsilon\over 2}\big)\,\,\big]^7
\Gamma(1 + 3 \epsilon) \over
\big(\,{1\over 2} -  {3\over 2}\,\mbox{\small$\epsilon$}\,\big)
\big(\,{1\over 2} +  {3\over 2}\,\mbox{\small$\epsilon$}\,\big)
\big(\,{1\over 2} -  {7\over 2}\,\mbox{\small$\epsilon$}\,\big)
\Gamma\big(\,{1\over 2} -  {7\over 2}\,\mbox{\small$\epsilon$}\,\big)
}\,.
\eea
Along with lower-range diagrams ${\cal A}_0, {\cal A}_1$ in (\ref{a0}, \ref{a1}), 
this contributes to the 6-loop part, ${\cal W}_6$, of $\ln  {\cal I }_{\rm bare} \,(\epsilon)$:
\bea
{\cal W}_6 (\epsilon)= \Big[{\cal B}_2 - 2 \cdot {1 \over 2} {\cal A}_0 {\cal A}_1 + {1 \over 3!} {\cal A}_0^3 \Big]^{\rm bare}
\cdot 
 4^{3}\,\,\, \mathbb{O}^{\rm in}_{6,6}\, .
\eea
We relegate details of computation of the operator content to section 6 and focus on the
coefficient of $\mathbb{O}^{\rm in}_{6,6}$. By multiplicative renormalizability, the leading singularity of ${\cal W}_6$ must begin at order of $O(\epsilon^{-1})$, viz. $\lim_{\epsilon\rightarrow 0}\,\,
\epsilon\, \ln \, {\cal I }_{\rm bare} \,(\epsilon)$ must be finite.
%
%
For this diagram, we found 
\bea
{\cal W}_6 (\epsilon)=
\Big[-{1\over 16}\,\, \epsilon^{-3}
-{25 + 9\,\psi(1)\over 48 }\,\, \epsilon^{-2}+O\big(\epsilon^{-1}\big)
\Big]
\cdot  4^{3}\,\, \mathbb{O}^{\rm in}_{6,6}\,.
\eea
%

Alternatively, we may opt to inject momentum $q$ at the middle skeleton vertex. In this case, we find
\bea
{\cal B}^{' \rm bare}_2 &=& (- (2 \pi)^2)^3 (-4)^3
C_3^2 
G(1+ \epsilon, 1)
G(2 + \epsilon - \omega, 1+ \epsilon)
 {1 \over (4 \pi)^{2 \omega}}
{1 \over (q^2)^{3 \epsilon}}
\nonumber \\
&=&
 {1 \over (4 \pi)^3}\,\, \Big[ {4 \pi \over q^2}
\Big]^{3 \epsilon}\,\,
{1\over 3\,\, \epsilon^3}\,\,
{\big[\,\,\Gamma \big({1\over 2}- {\epsilon\over 2}\big)\,\,\big]^7
\Gamma(1 + 3 \epsilon) \over
\big(\,{1\over 2} -  {3\over 2}\,\mbox{\small$\epsilon$}\,\big)^2
\big(\,{1\over 2} -  {7\over 2}\,\mbox{\small$\epsilon$}\,\big)\,\,
\Gamma\big(\,{1\over 2} -  {7\over 2}\,\mbox{\small$\epsilon$}\,\big)
}\,.
\eea
Hence the 6-loop part of the logarithm becomes
\bea
{\cal W}^{\ '}_6 (\epsilon) &= & \Big[{\cal B}'_2 - {\cal A}_0 {\cal A}_1  + {1 \over 6} {\cal A}^3_0
\Big]^{\rm bare}
\cdot 4^{3}\,\,\, \mathbb{O}^{\rm in}_{6,6}
\nonumber\\
&=& \Big[\,\, 0 \cdot \epsilon^{-3} +
{1\over 6 }\,\, \epsilon^{-2}+O\big(\epsilon^{-1}\big)\,\,
\Big]
\cdot
 4^{3}\,\,\, \mathbb{O}^{\rm in}_{6,6}\,.
\eea
%

The Feynman diagram ought to be infrared finite once independent momenta are injected to every external vertices. Unfortunately, the loop integral in this case is too complicated and do not permit all-loop
resummation. So, we see that in both options the problem stems from non-analyticity as some of
the injected momenta are taken to zero. 

\section{Proof the operator identities (\ref{identity66}) and (\ref{identityt66})}
To prove these two identities, we use the identity in (\ref{identity}) repeatedly.
For (\ref{identity66}), we first show the following identity:
\be
\mathbb{P}_{17} \mathbb{P}_{35}- \mathbb{P}_{15}\mathbb{P}_{37}
=\mathbb{P}_{13}\mathbb{P}_{35}\mathbb{P}_{57} +\mathbb{P}_{57}\mathbb{P}_{35}
\mathbb{P}_{13} -\mathbb{P}_{13} \mathbb{P}_{57} -\mathbb{P}_{15} -\mathbb{P}_{37}
+\mathbb{I}\,.
\ee
To show this, note that
\bea
&& \mathbb{P}_{17} \mathbb{P}_{35}-\mathbb{P}_{15}\mathbb{P}_{37}
= \Big(\mathbb{P}_{15} \mathbb{P}_{57} +\mathbb{P}_{57} \mathbb{P}_{15}
-\mathbb{P}_{15}- \mathbb{P}_{57} + \mathbb{I}
  \Big)  \mathbb{P}_{35}
\nonumber\\ &&\ \ \ \ \ \ \
- \mathbb{P}_{15} \Big( \mathbb{P}_{35} \mathbb{P}_{57}
 +\mathbb{P}_{57} \mathbb{P}_{35}
-\mathbb{P}_{35}- \mathbb{P}_{57} + \mathbb{I}  \Big) \nonumber\\
&&\ \ \ =
-\mathbb{P}_{35}\mathbb{P}_{13}\mathbb{P}_{57} +\mathbb{P}_{57}\mathbb{P}_{35}
\mathbb{P}_{13} -\mathbb{P}_{57}\mathbb{P}_{35}
-\mathbb{P}_{15}+\mathbb{P}_{35}+\mathbb{P}_{15} \mathbb{P}_{57} \nonumber\\
&&\ \ \
= -\mathbb{P}_{35}\mathbb{P}_{13}\mathbb{P}_{57} +\mathbb{P}_{57}\mathbb{P}_{35}
\mathbb{P}_{13} -\mathbb{P}_{57}\mathbb{P}_{35}
-\mathbb{P}_{15}+\mathbb{P}_{35}\nonumber\\
&&\ \ \ \ \ \ \ +
\Big( \mathbb{P}_{13} \mathbb{P}_{35}
 +\mathbb{P}_{35} \mathbb{P}_{13}
-\mathbb{P}_{13}- \mathbb{P}_{35} + \mathbb{I}  \Big)
\mathbb{P}_{57} \nonumber\\
&&\ \ \ =\mathbb{P}_{13}\mathbb{P}_{35}\mathbb{P}_{57} +\mathbb{P}_{57}\mathbb{P}_{35}
\mathbb{P}_{13} -\mathbb{P}_{13} \mathbb{P}_{57} -\mathbb{P}_{15} -\mathbb{P}_{37}
+\mathbb{I}\nonumber\\
&&\ \ \ \ \ \ \ -
\Big( \mathbb{P}_{35} \mathbb{P}_{57}
 +\mathbb{P}_{57} \mathbb{P}_{35}
-\mathbb{P}_{35}- \mathbb{P}_{57}-\mathbb{P}_{37} + \mathbb{I}  \Big)
\nonumber\\
&&\ \ \ =\mathbb{P}_{13}\mathbb{P}_{35}\mathbb{P}_{57} +\mathbb{P}_{57}\mathbb{P}_{35}
\mathbb{P}_{13} -\mathbb{P}_{13} \mathbb{P}_{57} -\mathbb{P}_{15} -\mathbb{P}_{37}
+\mathbb{I}\,,
\eea
Here, we used (\ref{identity}) to replace
$\mathbb{P}_{17}$ and $\mathbb{P}_{37}$ in the first line and
the second $\mathbb{P}_{15}$ in the third line.

Next, we turn to the operator $\mathbb{O}_{1\sim 7}$:
\bea
&& -2^{-3}\,\,\Big[\mathbb{O}_{123}\mathbb{O}_{345}\mathbb{O}_{567}+
\mathbb{O}_{567}\mathbb{O}_{345}\mathbb{O}_{123}\Big]\nonumber\\
&& = (\,2\,\mathbb{P}_{13}-\mathbb{I}\,)\, (\,2\,\mathbb{P}_{35}-\mathbb{I}\,)\,
(\,2\,\mathbb{P}_{57}-\mathbb{I}\,)+
 (\,2\,\mathbb{P}_{57}-\mathbb{I}\,)\, (\,2\,\mathbb{P}_{35}-\mathbb{I}\,)\,
(\,2\,\mathbb{P}_{13}-\mathbb{I}\,)\nonumber\\
&& =8\Big( \mathbb{P}_{13}\mathbb{P}_{35}\mathbb{P}_{57}
+\mathbb{P}_{57}\mathbb{P}_{35}
\mathbb{P}_{13} +\mathbb{P}_{13} \mathbb{P}_{57} -\mathbb{P}_{15} -\mathbb{P}_{37}
+\mathbb{I}
\Big)\nonumber\\
&& \ \ \ +4\mathbb{P}_{15}+ 4 \mathbb{P}_{37}-4 \mathbb{P}_{35}-2 \mathbb{I}
\nonumber\\
&&= 8\Big(
\mathbb{P}_{17} \mathbb{P}_{35}-\mathbb{P}_{15}\mathbb{P}_{37}
\Big)+4\mathbb{P}_{15}+ 4\mathbb{P}_{37}-4 \mathbb{P}_{35}-2 \mathbb{I}
\nonumber\\
&& =8\,\Big[\,
-64\mathbb{O}_{6,6}+80 \mathbb{O}_{4,4} -22\mathbb{O}_{2,2}+{1\over 4}\mathbb{I}
\,\Big]\,\,,
\eea
which is the proof of the identity (\ref{identity66}).

Finally, we turn to the identity in (\ref{identityt66}):
\bea
&& \mathbb{O}_{1\sim 7}+\widetilde{\mathbb{O}}_{1\sim 7}
= \mathbb{O}_{123}\,\Big(\,\mathbb{O}_{345}\mathbb{O}_{567}
+ \mathbb{O}_{567}\mathbb{O}_{345}\,\Big) +
\,\Big(\,\mathbb{O}_{345}\mathbb{O}_{567}
+ \mathbb{O}_{567}\mathbb{O}_{345}\,\Big)\,\mathbb{O}_{123}\nonumber\\
&&\ \ \ =2\,(\, \mathbb{I}-
2\,\mathbb{P}_{13}\,)\,(-8)\, (\, \mathbb{I}-
2\,\mathbb{P}_{37}\,)+(-8)\,(\, \mathbb{I}-
2\,\mathbb{P}_{37}\,)\, 2\, (\, \mathbb{I}-
2\,\mathbb{P}_{13}\,)\nonumber\\
&&\ \ \  = 32\, \Big(\,
\mathbb{I}-
2\,\mathbb{P}_{17}
\,\Big)\,.
\eea

\section{Derivation of $a_6(\epsilon)$ and $\tilde{a}_6(\epsilon)$}

The diagram in (\ref{fig1}) involves four 1-bubble and two 2-bubble propagators in total. The corresponding
factors are $(C_2)^4 (C_3)^2$. This then leads to the skeleton 4-loop
diagram in Fig.~\ref{skeleton}. We assign momentum and the $w$-factor
to each internal
line as follows; $k_1$ and $w_1=2-\omega$
for the upper-right line, $p-k_1-k_2$  and $w_5=2-\omega$  for
the middle-right line,
$k_2$  and $w_2=3-2\omega$
for the lower-right line, $k_3-k_1$  and $w_7=1$
for the upper-center line,
$k_4-k_2$  and $w_8=1$
for the lower-center line,
$k_3$  and $w_3=3-2\omega$
 for the upper-left line, $p-k_3-k_4$  and $w_6=2-\omega$
for the middle-left line,
 $k_4$  and $w_4=2-\omega$  for the lower-left line.
One has then
\be
a_6(\epsilon)\, (I_\epsilon)^7 (x^2\pi)^{9-6\omega}
=S \cdot
{1\over (4\pi)^{4\omega}} \cdot J_4(\epsilon)\cdot (C_2)^4 (C_3)^2
 \int {\rmd^{2\omega} p\over (2\pi)^{2\omega}}
{1\over (p^2)^{16-12\omega}} \, \, e^{i p \cdot x}\,,
\ee
where $S= (-4\pi^2)^3$ is the vertex and  symmetry factors.

For the Fourier transform, we used the formula,
\be
 \int {\rmd^{2\omega} p\over (2\pi)^{2\omega}}
{1\over (p^2)^\alpha} \, \, e^{i p \cdot x} =
{\Gamma(\omega-\alpha)\over 4^\alpha\pi^\omega \Gamma(\alpha)
}\,\, {1\over (x^2)^{\omega-\alpha}}
\,.
\ee
This leads to the expression $a_6(\epsilon)$ in (\ref{a6}).

The diagram in (\ref{fig3}) involves two
1-bubble  and three 2-bubble propagators in total.
This  leads to the skeleton 4-loop
diagram in Fig.~\ref{skeleton}. We assign
momentum and their $w$-factor
to each internal line as follows; $k_1$ and $w_1=2-\omega$
for the upper-right line, $p-k_1-k_2$  and $w_5=3-2\omega$  for
the middle-right line,
$k_2$  and $w_2=2-\omega$
for the lower-right line, $k_3-k_1$  and $w_7=1$
for the upper-center line,
$k_4-k_2$  and $w_8=1$
for the lower-center line,
$k_3$  and $w_3=3-2\omega$
 for the upper-left line, $p-k_3-k_4$  and $w_6=1$
for the middle-left line,
 $k_4$  and $w_4=3-2\omega$  for the lower-left line.
One has then
\be
\tilde{a}_6(\epsilon)\, (I_\epsilon)^7 (x^2\pi)^{9-6\omega}
=S \cdot
{1\over (4\pi)^{4\omega}} \cdot J_5(\epsilon)\cdot (C_2)^2 (C_3)^3
 \int {\rmd^{2\omega} p\over (2\pi)^{2\omega}}
{1\over (p^2)^{16-12\omega}} \, \, e^{i p \cdot x}\,.
\ee
This leads to the expression $a_6(\epsilon)$ in (\ref{ta6}).

\section{Evaluation of the integrals $J_1(\epsilon)$,
 $J_4(\epsilon)$ and $J_5(\epsilon)$}
	In this appendix, we evaluate the higher order terms
 of the two-loop integral $J_1(\epsilon)$
and two four-loop integrals,
 $J_4(\epsilon)$ and $J_5(\epsilon)$ in the section 6.
It is difficult to find their analytic forms in a direct manner.
We first numerically evaluate the integrals and then find the
corresponding expressions consisting of transcendental numbers.
There exists PSLQ algorithm\cite{pslq} which is quite useful in
finding such an analytic form from numerical data.
For the numerical evaluation, we use the Mathematica
packages--MB\cite{Czakon:2005rk,Smirnov:2009up} and
AMBRE\cite{Gluza:2007rt}. To find the corresponding analytic
expression, we use the package PSLQ.nb\cite{Bertok}, which is a
Mathematica implementation of the PSLQ algorithm.

Since these methods are based on the Mellin-Barnes representation
of Feynman integrals, we cast the integrals into the form:
\begin{eqnarray}
K_5(w_1,w_2,w_3,w_4,w_5)=\int\frac{\rmd z_1}
{2\pi i}\int\frac{\rmd z_2}{2\pi i}\frac{\Gamma(-z1)\,\,
\Gamma(\omega-w_{25}-z_1)\,\Gamma(\omega-w_{1}+z_1)}{\Gamma(w_1-z_1)}
\\ \nonumber
\otimes\,\,\frac{\Gamma(-z_2)\,\Gamma(\omega-w_{35}-z_2)\,
\Gamma(\omega-w_4+z_2)}{\Gamma(w_4-z_2)}\\  \nonumber
\otimes\,\, \frac{\Gamma(-\omega+w_{14}-z_{12})\,
\Gamma(-\omega+w_{235}+z_1+z_2)\,\Gamma(w_{5}+z_{12})}
{\Gamma(w_2)\Gamma(w_3)\,\Gamma(w_5)\,\Gamma(w_{235})\,
\Gamma(2\omega-w_{235})\,\Gamma(2\omega-w_{14}+z_{12})}\,.
\end{eqnarray}
with $w_{12\cdots}=w_1+w_2+\cdots$ and 
$z_{12\cdots}=z_1+z_1+\cdots$.
Direct application of these packages yields the following result:
\begin{eqnarray}
J_1(\epsilon)=K_5(2-\omega,3-2\omega,2-\omega,3-2\omega,1)=
\frac{\alpha^{-1}_{J_1}}{\epsilon}\,
\Big[\,1+\alpha^0_{J_1}\epsilon+\alpha^1_{J_2}
\epsilon^2+O(\epsilon^3)\,\Big],
\end{eqnarray}
where we have found  $\alpha^{-1}_{J_1}=-\frac{2}{3\pi }$
and $\alpha^0_{J_1}=13+\psi(\mbox{\small $\frac{1}{2}$})$ as reported in
\cite{BakMinRey}.
The numerical value of $\alpha^1_{J_1}$ is
obtained as
\be
4\alpha^1_{J_1}=309.7165844821997330276736227359\ldots\,.
\ee
By the PSLQ algorithm, we matched this numerical result as a linear combination
of $1$, $\pi^2$, $\psi(\mbox{\small $\frac{1}{2}$})$ and
 $\psi^2(\mbox{\small $\frac{1}{2}$})$ under the condition that their coefficients are rational
numbers. This basis follows from requiring that $e_{4,4}$ is a rational
number. This leads to the unique set of fractional coefficients:
\be
4\alpha^1_{J_1}= 488-\frac{17}{2}\,\,\pi^2+ 52\psi(\mbox{\small $\frac{1}{2}$}) +
2 \psi^2(\mbox{\small $\frac{1}{2}$})\,.
\ee
Note that the number of significant digits in this
numerical value is 30 and the difference between
these two expression is less than $10^{-30}$. We have checked
the other numbers, $\alpha^{-1}_{J_1}$ and $\alpha^0_{J_1}$
to the same accuracy.

The four-loop integrals $K_8(w_1,\cdots,w_8)$ in (\ref{qloop-2}) can be
evaluated in a similar way. Its Mellin-Barnes representation
 is given as
\begin{equation}
I_8(w_1,\cdots,w_8)=\int \frac{\rmd z_1}{2\pi i}\cdots\int
\frac{\rmd z_8}{2\pi i}\,\,\, \frac{N} {D}
\end{equation}
where
\begin{eqnarray}
N &=&\Big[\,\,\prod_{n=1}^8 \Gamma(-z_n)\,\,\Big]
\,\Gamma(w_{3}+z_{246})\,\Gamma(w_{7}+z_{235})\,
\Gamma({\omega}/{2}+z_{38})\,
\Gamma(\omega/2-w_{48}-z_1)
\nonumber \\
&&\otimes \,\,\Gamma({\omega}/{2}-
w_{6}-z_1)\,\Gamma(\omega-w_{3467}-z_{234})\,
\Gamma(\omega-w_{3678}-z_{256})   \nonumber \\
&&\otimes\,
\, \Gamma({3\omega}/{2}-
w_{3346778}-z_{1223456}) \,
\Gamma({3\omega}/{2}-w_{234678}-z_{23457})\,
 \nonumber \\
&&\otimes\,\,
\Gamma(-2\omega+w_{12345678}+z_{47})
\,
\Gamma({3\omega}/{2}-w_{2345678}-z_{23468})\,
\nonumber \\
&&\otimes\,\,
\Gamma(2\omega-w_{12345678}-z_{3478})\,
\Gamma(-{3\omega}/{2}+w_{2345678}+z_{23478})
 \nonumber \\
&&\otimes\,\,
\Gamma(-\omega+w_{234678}+z_{2345678})\,
\Gamma(-\omega+w_{34678}+z_{123456})
\end{eqnarray}
and
\begin{eqnarray}
D&=& \Gamma(w_3)\,\Gamma(w_5)\,\Gamma(w_6)\,\Gamma(w_7)\,\Gamma(w_8)\,
\Gamma(\omega-w_{468})\,\Gamma(w_5-z_5)\,\Gamma(-z_{38})
\nonumber \\
&&\otimes \,\,
\Gamma({3\omega}/{2}-w_{34678}-z_{1})\,\Gamma(2\omega-w_{2345678}-z_{234})
\nonumber \\
&&\otimes \,
\,\Gamma(2\omega-w_{33466778}-z_{223456})\,\Gamma(-\omega+w_{234678}+z_{23456})
\nonumber \\
&&\otimes\,\,
\Gamma({5\omega}/{2}-w_{12345678}-z_{47})\,
\Gamma(-{3\omega}/{2}+w_{12345678}+z_{3478})\,.
\end{eqnarray}

The $J_4(\epsilon)$  and $J_5(\epsilon)$, respectively, in (\ref{j4k})
and (\ref{j5k}) are defined with help of
$K_8$.
By numerical computations, we find that 
$J_4(\epsilon)$
has the expansion,
\begin{equation}
J_4(\epsilon)=
\frac{0.012008436579832\ldots}{\epsilon^2}\Big[ \,
1+ (50.0126265\ldots){\epsilon\over 2} + (360.321\ldots)\epsilon^2\,\Big].
\end{equation}
By the PSLQ algorithm, this can be converted to the following expression
\bea
J_4(\epsilon)&=&
\frac{16}{135\pi^2}\,\,\frac{1}{\epsilon^2}\,\,
\Big[ 1+ \Big(\,\frac{434}{15}+2\,\psi(\mbox{\small $\frac{1}{2}$})\Big) \,\epsilon
\nonumber\\
&&\  +
\Big(\,\frac{117301}{225}-\frac{67}{12}\pi^2+
\frac{868}{15}\psi(\mbox{\small $\frac{1}{2}$})+2\psi^2(\mbox{\small $\frac{1}{2}$})
\Big)\, \,\epsilon^2\,\,\Big]\,.
\label{c5}
\eea
The first two numbers can be calculated analytically
from the requirement of the multiplicative renormalizability. These agree with the
numerical values computed as above. In the last number, the coefficients of $\pi^2$,
$\psi(\mbox{\small $\frac{1}{2}$})$ and $\psi^2(\mbox{\small $\frac{1}{2}$})$
can be fixed by the assumption that $e_{6,6}$ is a fractional number.
Then the remaining fraction agrees with the numerical value
to six significant digits.

The last integral $J_5(\epsilon)$
is found to be 
\begin{equation}
J_5(\epsilon)=
\frac{0.11851851851\ldots}{\epsilon}\Big[\,\,
 1+ (21.8926\ldots)\,\epsilon + (571.233\ldots)\,{\epsilon^2\over 2}\,\,\Big]\,.
\end{equation}
This is equivalent to the following expression:
\begin{eqnarray}
&& J_5(\epsilon) = \frac{16}{135\epsilon}\,\,\Big[\, 1+
\Big(\,\frac{733}{30}+\psi(1)+\psi(
\mbox{\small $\frac{1}{2}$}
)\Big)\, \epsilon  \nonumber \\
&& \ \ +  \Big(\,\frac{177587}{225} - \frac{61}{6}\pi^2+
\frac{733}{15}\big[\,\psi(1)+\psi(\mbox{\small $\frac{1}{2}$})\big]+\big[\,\psi(1)+\psi(
\mbox{\small $\frac{1}{2}$}
)\,\big]^2 \, \Big)\,{\epsilon^2\over 2}\,\,\Big]\,.
\end{eqnarray}
We again checked the first two numbers from the renormalization of the underlying
 field theory. The last number is again determined in an analogous manner
to (\ref{c5}) and agrees with the numerical value up to six significant digits.

\end{document}